\documentclass[a4paper, traditabstract, longauth]{aa} 

%
\usepackage{graphicx}
\usepackage{color}
\usepackage{natbib}
\usepackage{longtable,lscape}
\usepackage{txfonts}
\usepackage{amsfonts}
\usepackage{amssymb}
\usepackage{newlfont}
\usepackage{textcomp}
\usepackage{times}
\usepackage{units}

\def\setsymbol#1#2{\expandafter\def\csname #1\endcsname{#2}}
\def\getsymbol#1{\csname #1\endcsname}

\def\Planck{{\it Planck\/}}



\newbox\tablebox    \newdimen\tablewidth
\def\leaderfil{\leaders\hbox to 5pt{\hss.\hss}\hfil}
%
%
\def\endPlancktable{\tablewidth=\columnwidth 
    $$\hss\copy\tablebox\hss$$
    \vskip-\lastskip\vskip -2pt}

\def\tablenote#1 #2\par{\begingroup \parindent=0.8em
    \abovedisplayshortskip=0pt\belowdisplayshortskip=0pt
    \noindent
    $$\hss\vbox{\hsize\tablewidth \hangindent=\parindent \hangafter=1 \noindent
    \hbox to \parindent{\sup{\rm #1}\hss}\strut#2\strut\par}\hss$$
    \endgroup}
\def\doubleline{\vskip 3pt\hrule \vskip 1.5pt \hrule \vskip 5pt}

%
\def\L2{\ifmmode L_2\else $L_2$\fi}

\def\DeltaT{\ifmmode \Delta T\else $\Delta T$\fi}
\def\deltat{\ifmmode \Delta t\else $\Delta t$\fi}
\def\fknee{\ifmmode f_{\rm knee}\else $f_{\rm knee}$\fi}
\def\Fmax{\ifmmode F_{\rm max}\else $F_{\rm max}$\fi}
\def\solar{\ifmmode{\rm M}_{\mathord\odot}\else${\rm M}_{\mathord\odot}$\fi}

\def\inv{\ifmmode^{-1}\else$^{-1}$\fi}
\def\mo{\ifmmode^{-1}\else$^{-1}$\fi}
\def\sup#1{\ifmmode ^{\rm #1}\else $^{\rm #1}$\fi}
\def\expo#1{\ifmmode \times 10^{#1}\else $\times 10^{#1}$\fi}
\def\,{\thinspace}
\def\lsim{\mathrel{\raise .4ex\hbox{\rlap{$<$}\lower 1.2ex\hbox{$\sim$}}}}
\def\gsim{\mathrel{\raise .4ex\hbox{\rlap{$>$}\lower 1.2ex\hbox{$\sim$}}}}

\def\simprop{\mathrel{\raise .4ex\hbox{\rlap{$\propto$}\lower 1.2ex\hbox{$\sim$}}}}
\def\deg{\ifmmode^\circ\else$^\circ$\fi}
\def\pdeg{\ifmmode $\setbox0=\hbox{$^{\circ}$}\rlap{\hskip.11\wd0 .}$^{\circ}
          \else \setbox0=\hbox{$^{\circ}$}\rlap{\hskip.11\wd0 .}$^{\circ}$\fi}
\def\arcs{\ifmmode {^{\scriptstyle\prime\prime}}
          \else $^{\scriptstyle\prime\prime}$\fi}
\def\arcm{\ifmmode {^{\scriptstyle\prime}}
          \else $^{\scriptstyle\prime}$\fi}
\newdimen\sa  \newdimen\sb
\def\parcs{\sa=.07em \sb=.03em
     \ifmmode \hbox{\rlap{.}}^{\scriptstyle\prime\kern -\sb\prime}\hbox{\kern -\sa}
     \else \rlap{.}$^{\scriptstyle\prime\kern -\sb\prime}$\kern -\sa\fi}
\def\parcm{\sa=.08em \sb=.03em
     \ifmmode \hbox{\rlap{.}\kern\sa}^{\scriptstyle\prime}\hbox{\kern-\sb}
     \else \rlap{.}\kern\sa$^{\scriptstyle\prime}$\kern-\sb\fi}
\def\ra[#1 #2 #3.#4]{#1\sup{h}#2\sup{m}#3\sup{s}\llap.#4}
\def\dec[#1 #2 #3.#4]{#1\deg#2\arcm#3\arcs\llap.#4}
\def\deco[#1 #2 #3]{#1\deg#2\arcm#3\arcs}
\def\rra[#1 #2]{#1\sup{h}#2\sup{m}}

\def\dots{\relax\ifmmode \ldots\else $\ldots$\fi}
%
%
\def\WHzsr{\ifmmode $W\,Hz\mo\,sr\mo$\else W\,Hz\mo\,sr\mo\fi}
\def\mHz{\ifmmode $\,mHz$\else \,mHz\fi}
\def\GHz{\ifmmode $\,GHz$\else \,GHz\fi}
\def\mKs{\ifmmode $\,mK\,s$^{1/2}\else \,mK\,s$^{1/2}$\fi}
\def\muKs{\ifmmode \,\mu$K\,s$^{1/2}\else \,$\mu$K\,s$^{1/2}$\fi}
\def\muKRJs{\ifmmode \,\mu$K$_{\rm RJ}$\,s$^{1/2}\else \,$\mu$K$_{\rm RJ}$\,s$^{1/2}$\fi}
\def\muKHz{\ifmmode \,\mu$K\,Hz$^{-1/2}\else \,$\mu$K\,Hz$^{-1/2}$\fi}
\def\MJysr{\ifmmode \,$MJy\,sr\mo$\else \,MJy\,sr\mo\fi}
\def\MJysrmK{\ifmmode \,$MJy\,sr\mo$\,mK$_{\rm CMB}\mo\else \,MJy\,sr\mo\,mK$_{\rm CMB}\mo$\fi}
\def\microns{\ifmmode \,\mu$m$\else \,$\mu$m\fi}

\def\muK{\ifmmode \,\mu$K$\else \,$\mu$\hbox{K}\fi}
\def\microK{\ifmmode \,\mu$K$\else \,$\mu$\hbox{K}\fi}
\def\muW{\ifmmode \,\mu$W$\else \,$\mu$\hbox{W}\fi}
\def\kms{\ifmmode $\,km\,s$^{-1}\else \,km\,s$^{-1}$\fi}
\def\kmsMpc{\ifmmode $\,\kms\,Mpc\mo$\else \,\kms\,Mpc\mo\fi}
%
%


\setsymbol{LFI:center:frequency:70GHz:units}{70.3\,GHz}
\setsymbol{LFI:center:frequency:44GHz:units}{44.1\,GHz}
\setsymbol{LFI:center:frequency:30GHz:units}{28.5\,GHz}

\setsymbol{LFI:center:frequency:70GHz}{70.3}
\setsymbol{LFI:center:frequency:44GHz}{44.1}
\setsymbol{LFI:center:frequency:30GHz}{28.5}

\setsymbol{LFI:center:frequency:LFI18:Rad:M:units}{71.7\GHz}
\setsymbol{LFI:center:frequency:LFI19:Rad:M:units}{67.5\GHz}
\setsymbol{LFI:center:frequency:LFI20:Rad:M:units}{69.2\GHz}
\setsymbol{LFI:center:frequency:LFI21:Rad:M:units}{70.4\GHz}
\setsymbol{LFI:center:frequency:LFI22:Rad:M:units}{71.5\GHz}
\setsymbol{LFI:center:frequency:LFI23:Rad:M:units}{70.8\GHz}
\setsymbol{LFI:center:frequency:LFI24:Rad:M:units}{44.4\GHz}
\setsymbol{LFI:center:frequency:LFI25:Rad:M:units}{44.0\GHz}
\setsymbol{LFI:center:frequency:LFI26:Rad:M:units}{43.9\GHz}
\setsymbol{LFI:center:frequency:LFI27:Rad:M:units}{28.3\GHz}
\setsymbol{LFI:center:frequency:LFI28:Rad:M:units}{28.8\GHz}
\setsymbol{LFI:center:frequency:LFI18:Rad:S:units}{70.1\GHz}
\setsymbol{LFI:center:frequency:LFI19:Rad:S:units}{69.6\GHz}
\setsymbol{LFI:center:frequency:LFI20:Rad:S:units}{69.5\GHz}
\setsymbol{LFI:center:frequency:LFI21:Rad:S:units}{69.5\GHz}
\setsymbol{LFI:center:frequency:LFI22:Rad:S:units}{72.8\GHz}
\setsymbol{LFI:center:frequency:LFI23:Rad:S:units}{71.3\GHz}
\setsymbol{LFI:center:frequency:LFI24:Rad:S:units}{44.1\GHz}
\setsymbol{LFI:center:frequency:LFI25:Rad:S:units}{44.1\GHz}
\setsymbol{LFI:center:frequency:LFI26:Rad:S:units}{44.1\GHz}
\setsymbol{LFI:center:frequency:LFI27:Rad:S:units}{28.5\GHz}
\setsymbol{LFI:center:frequency:LFI28:Rad:S:units}{28.2\GHz}

\setsymbol{LFI:center:frequency:LFI18:Rad:M}{71.7}
\setsymbol{LFI:center:frequency:LFI19:Rad:M}{67.5}
\setsymbol{LFI:center:frequency:LFI20:Rad:M}{69.2}
\setsymbol{LFI:center:frequency:LFI21:Rad:M}{70.4}
\setsymbol{LFI:center:frequency:LFI22:Rad:M}{71.5}
\setsymbol{LFI:center:frequency:LFI23:Rad:M}{70.8}
\setsymbol{LFI:center:frequency:LFI24:Rad:M}{44.4}
\setsymbol{LFI:center:frequency:LFI25:Rad:M}{44.0}
\setsymbol{LFI:center:frequency:LFI26:Rad:M}{43.9}
\setsymbol{LFI:center:frequency:LFI27:Rad:M}{28.3}
\setsymbol{LFI:center:frequency:LFI28:Rad:M}{28.8}
\setsymbol{LFI:center:frequency:LFI18:Rad:S}{70.1}
\setsymbol{LFI:center:frequency:LFI19:Rad:S}{69.6}
\setsymbol{LFI:center:frequency:LFI20:Rad:S}{69.5}
\setsymbol{LFI:center:frequency:LFI21:Rad:S}{69.5}
\setsymbol{LFI:center:frequency:LFI22:Rad:S}{72.8}
\setsymbol{LFI:center:frequency:LFI23:Rad:S}{71.3}
\setsymbol{LFI:center:frequency:LFI24:Rad:S}{44.1}
\setsymbol{LFI:center:frequency:LFI25:Rad:S}{44.1}
\setsymbol{LFI:center:frequency:LFI26:Rad:S}{44.1}
\setsymbol{LFI:center:frequency:LFI27:Rad:S}{28.5}
\setsymbol{LFI:center:frequency:LFI28:Rad:S}{28.2}


\setsymbol{LFI:white:noise:sensitivity:70GHz:units}{134.7\muKs}
\setsymbol{LFI:white:noise:sensitivity:44GHz:units}{164.7\muKs}
\setsymbol{LFI:white:noise:sensitivity:30GHz:units}{143.4\muKs}

\setsymbol{LFI:white:noise:sensitivity:70GHz}{134.7}
\setsymbol{LFI:white:noise:sensitivity:44GHz}{164.7}
\setsymbol{LFI:white:noise:sensitivity:30GHz}{143.4}


\setsymbol{LFI:white:noise:sensitivity:LFI18:Rad:M:units}{512.0\muKs}
\setsymbol{LFI:white:noise:sensitivity:LFI19:Rad:M:units}{581.4\muKs}
\setsymbol{LFI:white:noise:sensitivity:LFI20:Rad:M:units}{590.8\muKs}
\setsymbol{LFI:white:noise:sensitivity:LFI21:Rad:M:units}{455.2\muKs}
\setsymbol{LFI:white:noise:sensitivity:LFI22:Rad:M:units}{492.0\muKs}
\setsymbol{LFI:white:noise:sensitivity:LFI23:Rad:M:units}{507.7\muKs}
\setsymbol{LFI:white:noise:sensitivity:LFI24:Rad:M:units}{462.2\muKs}
\setsymbol{LFI:white:noise:sensitivity:LFI25:Rad:M:units}{413.6\muKs}
\setsymbol{LFI:white:noise:sensitivity:LFI26:Rad:M:units}{478.6\muKs}
\setsymbol{LFI:white:noise:sensitivity:LFI27:Rad:M:units}{277.7\muKs}
\setsymbol{LFI:white:noise:sensitivity:LFI28:Rad:M:units}{312.3\muKs}
\setsymbol{LFI:white:noise:sensitivity:LFI18:Rad:S:units}{465.7\muKs}
\setsymbol{LFI:white:noise:sensitivity:LFI19:Rad:S:units}{555.6\muKs}
\setsymbol{LFI:white:noise:sensitivity:LFI20:Rad:S:units}{623.2\muKs}
\setsymbol{LFI:white:noise:sensitivity:LFI21:Rad:S:units}{564.1\muKs}
\setsymbol{LFI:white:noise:sensitivity:LFI22:Rad:S:units}{534.4\muKs}
\setsymbol{LFI:white:noise:sensitivity:LFI23:Rad:S:units}{542.4\muKs}
\setsymbol{LFI:white:noise:sensitivity:LFI24:Rad:S:units}{399.2\muKs}
\setsymbol{LFI:white:noise:sensitivity:LFI25:Rad:S:units}{392.6\muKs}
\setsymbol{LFI:white:noise:sensitivity:LFI26:Rad:S:units}{418.6\muKs}
\setsymbol{LFI:white:noise:sensitivity:LFI27:Rad:S:units}{302.9\muKs}
\setsymbol{LFI:white:noise:sensitivity:LFI28:Rad:S:units}{285.3\muKs}

\setsymbol{LFI:white:noise:sensitivity:LFI18:Rad:M}{512.0}
\setsymbol{LFI:white:noise:sensitivity:LFI19:Rad:M}{581.4}
\setsymbol{LFI:white:noise:sensitivity:LFI20:Rad:M}{590.8}
\setsymbol{LFI:white:noise:sensitivity:LFI21:Rad:M}{455.2}
\setsymbol{LFI:white:noise:sensitivity:LFI22:Rad:M}{492.0}
\setsymbol{LFI:white:noise:sensitivity:LFI23:Rad:M}{507.7}
\setsymbol{LFI:white:noise:sensitivity:LFI24:Rad:M}{462.2}
\setsymbol{LFI:white:noise:sensitivity:LFI25:Rad:M}{413.6}
\setsymbol{LFI:white:noise:sensitivity:LFI26:Rad:M}{478.6}
\setsymbol{LFI:white:noise:sensitivity:LFI27:Rad:M}{277.7}
\setsymbol{LFI:white:noise:sensitivity:LFI28:Rad:M}{312.3}
\setsymbol{LFI:white:noise:sensitivity:LFI18:Rad:S}{465.7}
\setsymbol{LFI:white:noise:sensitivity:LFI19:Rad:S}{555.6}
\setsymbol{LFI:white:noise:sensitivity:LFI20:Rad:S}{623.2}
\setsymbol{LFI:white:noise:sensitivity:LFI21:Rad:S}{564.1}
\setsymbol{LFI:white:noise:sensitivity:LFI22:Rad:S}{534.4}
\setsymbol{LFI:white:noise:sensitivity:LFI23:Rad:S}{542.4}
\setsymbol{LFI:white:noise:sensitivity:LFI24:Rad:S}{399.2}
\setsymbol{LFI:white:noise:sensitivity:LFI25:Rad:S}{392.6}
\setsymbol{LFI:white:noise:sensitivity:LFI26:Rad:S}{418.6}
\setsymbol{LFI:white:noise:sensitivity:LFI27:Rad:S}{302.9}
\setsymbol{LFI:white:noise:sensitivity:LFI28:Rad:S}{285.3}


\setsymbol{LFI:knee:frequency:70GHz:units}{29.5\mHz}
\setsymbol{LFI:knee:frequency:44GHz:units}{56.2\mHz}
\setsymbol{LFI:knee:frequency:30GHz:units}{113.7\mHz}

\setsymbol{LFI:knee:frequency:70GHz}{29.5}
\setsymbol{LFI:knee:frequency:44GHz}{56.2}
\setsymbol{LFI:knee:frequency:30GHz}{113.7}

\setsymbol{LFI:knee:frequency:LFI18:Rad:M:units}{16.3\mHz}
\setsymbol{LFI:knee:frequency:LFI19:Rad:M:units}{15.1\mHz}
\setsymbol{LFI:knee:frequency:LFI20:Rad:M:units}{18.7\mHz}
\setsymbol{LFI:knee:frequency:LFI21:Rad:M:units}{37.2\mHz}
\setsymbol{LFI:knee:frequency:LFI22:Rad:M:units}{12.7\mHz}
\setsymbol{LFI:knee:frequency:LFI23:Rad:M:units}{34.6\mHz}
\setsymbol{LFI:knee:frequency:LFI24:Rad:M:units}{46.2\mHz}
\setsymbol{LFI:knee:frequency:LFI25:Rad:M:units}{24.9\mHz}
\setsymbol{LFI:knee:frequency:LFI26:Rad:M:units}{67.6\mHz}
\setsymbol{LFI:knee:frequency:LFI27:Rad:M:units}{187.4\mHz}
\setsymbol{LFI:knee:frequency:LFI28:Rad:M:units}{122.2\mHz}
\setsymbol{LFI:knee:frequency:LFI18:Rad:S:units}{17.7\mHz}
\setsymbol{LFI:knee:frequency:LFI19:Rad:S:units}{22.0\mHz}
\setsymbol{LFI:knee:frequency:LFI20:Rad:S:units}{8.7\mHz}
\setsymbol{LFI:knee:frequency:LFI21:Rad:S:units}{25.9\mHz}
\setsymbol{LFI:knee:frequency:LFI22:Rad:S:units}{15.8\mHz}
\setsymbol{LFI:knee:frequency:LFI23:Rad:S:units}{129.8\mHz}
\setsymbol{LFI:knee:frequency:LFI24:Rad:S:units}{100.9\mHz}
\setsymbol{LFI:knee:frequency:LFI25:Rad:S:units}{38.9\mHz}
\setsymbol{LFI:knee:frequency:LFI26:Rad:S:units}{58.9\mHz}
\setsymbol{LFI:knee:frequency:LFI27:Rad:S:units}{104.4\mHz}
\setsymbol{LFI:knee:frequency:LFI28:Rad:S:units}{40.7\mHz}

\setsymbol{LFI:knee:frequency:LFI18:Rad:M}{16.3}
\setsymbol{LFI:knee:frequency:LFI19:Rad:M}{15.1}
\setsymbol{LFI:knee:frequency:LFI20:Rad:M}{18.7}
\setsymbol{LFI:knee:frequency:LFI21:Rad:M}{37.2}
\setsymbol{LFI:knee:frequency:LFI22:Rad:M}{12.7}
\setsymbol{LFI:knee:frequency:LFI23:Rad:M}{34.6}
\setsymbol{LFI:knee:frequency:LFI24:Rad:M}{46.2}
\setsymbol{LFI:knee:frequency:LFI25:Rad:M}{24.9}
\setsymbol{LFI:knee:frequency:LFI26:Rad:M}{67.6}
\setsymbol{LFI:knee:frequency:LFI27:Rad:M}{187.4}
\setsymbol{LFI:knee:frequency:LFI28:Rad:M}{122.2}
\setsymbol{LFI:knee:frequency:LFI18:Rad:S}{17.7}
\setsymbol{LFI:knee:frequency:LFI19:Rad:S}{22.0}
\setsymbol{LFI:knee:frequency:LFI20:Rad:S}{8.7}
\setsymbol{LFI:knee:frequency:LFI21:Rad:S}{25.9}
\setsymbol{LFI:knee:frequency:LFI22:Rad:S}{15.8}
\setsymbol{LFI:knee:frequency:LFI23:Rad:S}{129.8}
\setsymbol{LFI:knee:frequency:LFI24:Rad:S}{100.9}
\setsymbol{LFI:knee:frequency:LFI25:Rad:S}{38.9}
\setsymbol{LFI:knee:frequency:LFI26:Rad:S}{58.9}
\setsymbol{LFI:knee:frequency:LFI27:Rad:S}{104.4}
\setsymbol{LFI:knee:frequency:LFI28:Rad:S}{40.7}


\setsymbol{LFI:slope:70GHz:units}{$-1.03$\mHz}
\setsymbol{LFI:slope:44GHz:units}{$-0.89$\mHz}
\setsymbol{LFI:slope:30GHz:units}{$-0.87$\mHz}

\setsymbol{LFI:slope:70GHz}{$-1.03$}
\setsymbol{LFI:slope:44GHz}{$-0.89$}
\setsymbol{LFI:slope:30GHz}{$-0.87$}

\setsymbol{LFI:slope:LFI18:Rad:M:units}{$-1.04$\mHz}
\setsymbol{LFI:slope:LFI19:Rad:M:units}{$-1.09$\mHz}
\setsymbol{LFI:slope:LFI20:Rad:M:units}{$-0.69$\mHz}
\setsymbol{LFI:slope:LFI21:Rad:M:units}{$-1.56$\mHz}
\setsymbol{LFI:slope:LFI22:Rad:M:units}{$-1.01$\mHz}
\setsymbol{LFI:slope:LFI23:Rad:M:units}{$-0.96$\mHz}
\setsymbol{LFI:slope:LFI24:Rad:M:units}{$-0.83$\mHz}
\setsymbol{LFI:slope:LFI25:Rad:M:units}{$-0.91$\mHz}
\setsymbol{LFI:slope:LFI26:Rad:M:units}{$-0.95$\mHz}
\setsymbol{LFI:slope:LFI27:Rad:M:units}{$-0.87$\mHz}
\setsymbol{LFI:slope:LFI28:Rad:M:units}{$-0.88$\mHz}
\setsymbol{LFI:slope:LFI18:Rad:S:units}{$-1.15$\mHz}
\setsymbol{LFI:slope:LFI19:Rad:S:units}{$-1.00$\mHz}
\setsymbol{LFI:slope:LFI20:Rad:S:units}{$-0.95$\mHz}
\setsymbol{LFI:slope:LFI21:Rad:S:units}{$-0.92$\mHz}
\setsymbol{LFI:slope:LFI22:Rad:S:units}{$-1.01$\mHz}
\setsymbol{LFI:slope:LFI23:Rad:S:units}{$-0.95$\mHz}
\setsymbol{LFI:slope:LFI24:Rad:S:units}{$-0.73$\mHz}
\setsymbol{LFI:slope:LFI25:Rad:S:units}{$-1.16$\mHz}
\setsymbol{LFI:slope:LFI26:Rad:S:units}{$-0.79$\mHz}
\setsymbol{LFI:slope:LFI27:Rad:S:units}{$-0.82$\mHz}
\setsymbol{LFI:slope:LFI28:Rad:S:units}{$-0.91$\mHz}

\setsymbol{LFI:slope:LFI18:Rad:M}{$-1.04$}
\setsymbol{LFI:slope:LFI19:Rad:M}{$-1.09$}
\setsymbol{LFI:slope:LFI20:Rad:M}{$-0.69$}
\setsymbol{LFI:slope:LFI21:Rad:M}{$-1.56$}
\setsymbol{LFI:slope:LFI22:Rad:M}{$-1.01$}
\setsymbol{LFI:slope:LFI23:Rad:M}{$-0.96$}
\setsymbol{LFI:slope:LFI24:Rad:M}{$-0.83$}
\setsymbol{LFI:slope:LFI25:Rad:M}{$-0.91$}
\setsymbol{LFI:slope:LFI26:Rad:M}{$-0.95$}
\setsymbol{LFI:slope:LFI27:Rad:M}{$-0.87$}
\setsymbol{LFI:slope:LFI28:Rad:M}{$-0.88$}
\setsymbol{LFI:slope:LFI18:Rad:S}{$-1.15$}
\setsymbol{LFI:slope:LFI19:Rad:S}{$-1.00$}
\setsymbol{LFI:slope:LFI20:Rad:S}{$-0.95$}
\setsymbol{LFI:slope:LFI21:Rad:S}{$-0.92$}
\setsymbol{LFI:slope:LFI22:Rad:S}{$-1.01$}
\setsymbol{LFI:slope:LFI23:Rad:S}{$-0.95$}
\setsymbol{LFI:slope:LFI24:Rad:S}{$-0.73$}
\setsymbol{LFI:slope:LFI25:Rad:S}{$-1.16$}
\setsymbol{LFI:slope:LFI26:Rad:S}{$-0.79$}
\setsymbol{LFI:slope:LFI27:Rad:S}{$-0.82$}
\setsymbol{LFI:slope:LFI28:Rad:S}{$-0.91$}


\setsymbol{LFI:FWHM:70GHz:units}{13\parcm01}
\setsymbol{LFI:FWHM:44GHz:units}{27\parcm92}
\setsymbol{LFI:FWHM:30GHz:units}{32\parcm65}

\setsymbol{LFI:FWHM:70GHz}{13.01}
\setsymbol{LFI:FWHM:44GHz}{27.92}
\setsymbol{LFI:FWHM:30GHz}{32.65}

\setsymbol{LFI:FWHM:LFI18:units}{13\parcm39}
\setsymbol{LFI:FWHM:LFI19:units}{13\parcm01}
\setsymbol{LFI:FWHM:LFI20:units}{12\parcm75}
\setsymbol{LFI:FWHM:LFI21:units}{12\parcm74}
\setsymbol{LFI:FWHM:LFI22:units}{12\parcm87}
\setsymbol{LFI:FWHM:LFI23:units}{13\parcm27}
\setsymbol{LFI:FWHM:LFI24:units}{22\parcm98}
\setsymbol{LFI:FWHM:LFI25:units}{30\parcm46}
\setsymbol{LFI:FWHM:LFI26:units}{30\parcm31}
\setsymbol{LFI:FWHM:LFI27:units}{32\parcm65}
\setsymbol{LFI:FWHM:LFI28:units}{32\parcm66}

\setsymbol{LFI:FWHM:LFI18}{13.39}
\setsymbol{LFI:FWHM:LFI19}{13.01}
\setsymbol{LFI:FWHM:LFI20}{12.75}
\setsymbol{LFI:FWHM:LFI21}{12.74}
\setsymbol{LFI:FWHM:LFI22}{12.87}
\setsymbol{LFI:FWHM:LFI23}{13.27}
\setsymbol{LFI:FWHM:LFI24}{22.98}
\setsymbol{LFI:FWHM:LFI25}{30.46}
\setsymbol{LFI:FWHM:LFI26}{30.31}
\setsymbol{LFI:FWHM:LFI27}{32.65}
\setsymbol{LFI:FWHM:LFI28}{32.66}



\setsymbol{LFI:FWHM:uncertainty:LFI18:units}{0.170\arcm}
\setsymbol{LFI:FWHM:uncertainty:LFI19:units}{0.174\arcm}
\setsymbol{LFI:FWHM:uncertainty:LFI20:units}{0.170\arcm}
\setsymbol{LFI:FWHM:uncertainty:LFI21:units}{0.156\arcm}
\setsymbol{LFI:FWHM:uncertainty:LFI22:units}{0.164\arcm}
\setsymbol{LFI:FWHM:uncertainty:LFI23:units}{0.171\arcm}
\setsymbol{LFI:FWHM:uncertainty:LFI24:units}{0.652\arcm}
\setsymbol{LFI:FWHM:uncertainty:LFI25:units}{1.075\arcm}
\setsymbol{LFI:FWHM:uncertainty:LFI26:units}{1.131\arcm}
\setsymbol{LFI:FWHM:uncertainty:LFI27:units}{1.266\arcm}
\setsymbol{LFI:FWHM:uncertainty:LFI28:units}{1.287\arcm}

\setsymbol{LFI:FWHM:uncertainty:LFI18}{0.170}
\setsymbol{LFI:FWHM:uncertainty:LFI19}{0.174}
\setsymbol{LFI:FWHM:uncertainty:LFI20}{0.170}
\setsymbol{LFI:FWHM:uncertainty:LFI21}{0.156}
\setsymbol{LFI:FWHM:uncertainty:LFI22}{0.164}
\setsymbol{LFI:FWHM:uncertainty:LFI23}{0.171}
\setsymbol{LFI:FWHM:uncertainty:LFI24}{0.652}
\setsymbol{LFI:FWHM:uncertainty:LFI25}{1.075}
\setsymbol{LFI:FWHM:uncertainty:LFI26}{1.131}
\setsymbol{LFI:FWHM:uncertainty:LFI27}{1.266}
\setsymbol{LFI:FWHM:uncertainty:LFI28}{1.287}


\setsymbol{HFI:center:frequency:100GHz:units}{100\,GHz}
\setsymbol{HFI:center:frequency:143GHz:units}{143\,GHz}
\setsymbol{HFI:center:frequency:217GHz:units}{217\,GHz}
\setsymbol{HFI:center:frequency:353GHz:units}{353\,GHz}
\setsymbol{HFI:center:frequency:545GHz:units}{545\,GHz}
\setsymbol{HFI:center:frequency:857GHz:units}{857\,GHz}

\setsymbol{HFI:center:frequency:100GHz}{100}
\setsymbol{HFI:center:frequency:143GHz}{143}
\setsymbol{HFI:center:frequency:217GHz}{217}
\setsymbol{HFI:center:frequency:353GHz}{353}
\setsymbol{HFI:center:frequency:545GHz}{545}
\setsymbol{HFI:center:frequency:857GHz}{857}


\setsymbol{HFI:Ndetectors:100GHz}{8}
\setsymbol{HFI:Ndetectors:143GHz}{11}
\setsymbol{HFI:Ndetectors:217GHz}{12}
\setsymbol{HFI:Ndetectors:353GHz}{12}
\setsymbol{HFI:Ndetectors:545GHz}{3}
\setsymbol{HFI:Ndetectors:857GHz}{4}


\setsymbol{HFI:FWHM:Maps:100GHz:units}{9\parcm88}
\setsymbol{HFI:FWHM:Maps:143GHz:units}{7\parcm18}
\setsymbol{HFI:FWHM:Maps:217GHz:units}{4\parcm87}
\setsymbol{HFI:FWHM:Maps:353GHz:units}{4\parcm65}
\setsymbol{HFI:FWHM:Maps:545GHz:units}{4\parcm72}
\setsymbol{HFI:FWHM:Maps:857GHz:units}{4\parcm39}
\setsymbol{HFI:FWHM:Maps:100GHz}{9.88}
\setsymbol{HFI:FWHM:Maps:143GHz}{7.18}
\setsymbol{HFI:FWHM:Maps:217GHz}{4.87}
\setsymbol{HFI:FWHM:Maps:353GHz}{4.65}
\setsymbol{HFI:FWHM:Maps:545GHz}{4.72}
\setsymbol{HFI:FWHM:Maps:857GHz}{4.39}


\setsymbol{HFI:beam:ellipticity:Maps:100GHz}{1.15}
\setsymbol{HFI:beam:ellipticity:Maps:143GHz}{1.01}
\setsymbol{HFI:beam:ellipticity:Maps:217GHz}{1.06}
\setsymbol{HFI:beam:ellipticity:Maps:353GHz}{1.05}
\setsymbol{HFI:beam:ellipticity:Maps:545GHz}{1.14}
\setsymbol{HFI:beam:ellipticity:Maps:857GHz}{1.19}


\setsymbol{HFI:FWHM:Mars:100GHz:units}{9\parcm37}
\setsymbol{HFI:FWHM:Mars:143GHz:units}{7\parcm04}
\setsymbol{HFI:FWHM:Mars:217GHz:units}{4\parcm68}
\setsymbol{HFI:FWHM:Mars:353GHz:units}{4\parcm43}
\setsymbol{HFI:FWHM:Mars:545GHz:units}{3\parcm80}
\setsymbol{HFI:FWHM:Mars:857GHz:units}{3\parcm67}

\setsymbol{HFI:FWHM:Mars:100GHz}{9.37}
\setsymbol{HFI:FWHM:Mars:143GHz}{7.04}
\setsymbol{HFI:FWHM:Mars:217GHz}{4.68}
\setsymbol{HFI:FWHM:Mars:353GHz}{4.43}
\setsymbol{HFI:FWHM:Mars:545GHz}{3.80}
\setsymbol{HFI:FWHM:Mars:857GHz}{3.67}


\setsymbol{HFI:beam:ellipticity:Mars:100GHz}{1.18}
\setsymbol{HFI:beam:ellipticity:Mars:143GHz}{1.03}
\setsymbol{HFI:beam:ellipticity:Mars:217GHz}{1.14}
\setsymbol{HFI:beam:ellipticity:Mars:353GHz}{1.09}
\setsymbol{HFI:beam:ellipticity:Mars:545GHz}{1.25}
\setsymbol{HFI:beam:ellipticity:Mars:857GHz}{1.03}


\setsymbol{HFI:CMB:relative:calibration:100GHz}{$\lsim 1\%$}
\setsymbol{HFI:CMB:relative:calibration:143GHz}{$\lsim 1\%$}
\setsymbol{HFI:CMB:relative:calibration:217GHz}{$\lsim 1\%$}
\setsymbol{HFI:CMB:relative:calibration:353GHz}{$\lsim 1\%$}
\setsymbol{HFI:CMB:relative:calibration:545GHz}{}
\setsymbol{HFI:CMB:relative:calibration:857GHz}{}


\setsymbol{HFI:CMB:absolute:calibration:100GHz}{$\lsim 2\%$}
\setsymbol{HFI:CMB:absolute:calibration:143GHz}{$\lsim 2\%$}
\setsymbol{HFI:CMB:absolute:calibration:217GHz}{$\lsim 2\%$}
\setsymbol{HFI:CMB:absolute:calibration:353GHz}{$\lsim 2\%$}
\setsymbol{HFI:CMB:absolute:calibration:545GHz}{}
\setsymbol{HFI:CMB:absolute:calibration:857GHz}{}


\setsymbol{HFI:FIRAS:gain:calibration:accuracy:statistical:100GHz}{}
\setsymbol{HFI:FIRAS:gain:calibration:accuracy:statistical:143GHz}{}
\setsymbol{HFI:FIRAS:gain:calibration:accuracy:statistical:217GHz}{}
\setsymbol{HFI:FIRAS:gain:calibration:accuracy:statistical:353GHz}{2.5\%}
\setsymbol{HFI:FIRAS:gain:calibration:accuracy:statistical:545GHz}{1\%}
\setsymbol{HFI:FIRAS:gain:calibration:accuracy:statistical:857GHz}{0.5\%}


\setsymbol{HFI:FIRAS:gain:calibration:accuracy:systematic:100GHz}{}
\setsymbol{HFI:FIRAS:gain:calibration:accuracy:systematic:143GHz}{}
\setsymbol{HFI:FIRAS:gain:calibration:accuracy:systematic:217GHz}{}
\setsymbol{HFI:FIRAS:gain:calibration:accuracy:systematic:353GHz}{}
\setsymbol{HFI:FIRAS:gain:calibration:accuracy:systematic:545GHz}{7\%}
\setsymbol{HFI:FIRAS:gain:calibration:accuracy:systematic:857GHz}{7\%}


\setsymbol{HFI:FIRAS:zero:point:accuracy:100GHz:units}{0.8\MJysr}
\setsymbol{HFI:FIRAS:zero:point:accuracy:143GHz:units}{}
\setsymbol{HFI:FIRAS:zero:point:accuracy:217GHz:units}{}
\setsymbol{HFI:FIRAS:zero:point:accuracy:353GHz:units}{1.4\MJysr}
\setsymbol{HFI:FIRAS:zero:point:accuracy:545GHz:units}{2.2\MJysr}
\setsymbol{HFI:FIRAS:zero:point:accuracy:857GHz:units}{1.7\MJysr}

\setsymbol{HFI:FIRAS:zero:point:accuracy:100GHz}{0.8}
\setsymbol{HFI:FIRAS:zero:point:accuracy:143GHz}{}
\setsymbol{HFI:FIRAS:zero:point:accuracy:217GHz}{}
\setsymbol{HFI:FIRAS:zero:point:accuracy:353GHz}{1.4}
\setsymbol{HFI:FIRAS:zero:point:accuracy:545GHz}{2.2}
\setsymbol{HFI:FIRAS:zero:point:accuracy:857GHz}{1.7}


\setsymbol{HFI:unit:conversion:100GHz:units}{0.2415\MJysrmK}
\setsymbol{HFI:unit:conversion:143GHz:units}{0.3694\MJysrmK}
\setsymbol{HFI:unit:conversion:217GHz:units}{0.4811\MJysrmK}
\setsymbol{HFI:unit:conversion:353GHz:units}{0.2883\MJysrmK}
\setsymbol{HFI:unit:conversion:545GHz:units}{0.05826\MJysrmK}
\setsymbol{HFI:unit:conversion:857GHz:units}{0.002238\MJysrmK}

\setsymbol{HFI:unit:conversion:100GHz}{0.2415}
\setsymbol{HFI:unit:conversion:143GHz}{0.3694}
\setsymbol{HFI:unit:conversion:217GHz}{0.4811}
\setsymbol{HFI:unit:conversion:353GHz}{0.2883}
\setsymbol{HFI:unit:conversion:545GHz}{0.05826}
\setsymbol{HFI:unit:conversion:857GHz}{0.002238}


\setsymbol{HFI:colour:correction:alpha=-2:V1.01:100GHz}{0.9893}
\setsymbol{HFI:colour:correction:alpha=-2:V1.01:143GHz}{0.9759}
\setsymbol{HFI:colour:correction:alpha=-2:V1.01:217GHz}{1.0007}
\setsymbol{HFI:colour:correction:alpha=-2:V1.01:353GHz}{1.0028}
\setsymbol{HFI:colour:correction:alpha=-2:V1.01:545GHz}{1.0019}
\setsymbol{HFI:colour:correction:alpha=-2:V1.01:857GHz}{0.9889}


\setsymbol{HFI:colour:correction:alpha=0:V1.01:100GHz}{1.0008}
\setsymbol{HFI:colour:correction:alpha=0:V1.01:143GHz}{1.0148}
\setsymbol{HFI:colour:correction:alpha=0:V1.01:217GHz}{0.9909}
\setsymbol{HFI:colour:correction:alpha=0:V1.01:353GHz}{0.9888}
\setsymbol{HFI:colour:correction:alpha=0:V1.01:545GHz}{0.9878}
\setsymbol{HFI:colour:correction:alpha=0:V1.01:857GHz}{1.0014}

\def\reff@jnl#1{{\rm#1\/}}
\def\apj{\reff@jnl{ApJ}}       
\def\apjs{\reff@jnl{ApJS}}     
\def\aaps{\reff@jnl{A\&AS}}    
\def\mnras{\reff@jnl{MNRAS}}   
\def\prd{\reff@jnl{Phys.\ Rev.\ D}}    


\newcommand{\beq}{\begin{equation}}
\newcommand{\eeq}{\end{equation}}

\newcommand{\be}{\begin{equation}}
\newcommand{\ee}{\end{equation}}
\newcommand{\bea}{\begin{eq}}
\newcommand{\eea}{\end{equation}}

{\begin{enumerate}\setlength{\itemsep}{0mm}}%
{\end{enumerate}}
{\begin{enumerate}\setlength{\itemsep}{0mm}}%
{\end{enumerate}}

\newcommand{\bc}{\begin{center}}
\newcommand{\ec}{\end{center}}
\newcommand{\bi}{\begin{itemize}}
\newcommand{\ei}{\end{itemize}}
\newcommand{\ben}{\begin{enumerate}}
\newcommand{\een}{\end{enumerate}}

\usepackage{mathrsfs}
\usepackage{mathbbol}

\newfont{\gwpfont}{cmssq8 scaled 1000}


\begin{document}

\title{\Planck\ 2013 results. XXI. All-sky Compton parameter power
 spectrum and high-order statistics}

\author{\small
Planck Collaboration:
P.~A.~R.~Ade\inst{89}
\and
N.~Aghanim\inst{63}
\and
C.~Armitage-Caplan\inst{95}
\and
M.~Arnaud\inst{76}
\and
M.~Ashdown\inst{73, 6}
\and
F.~Atrio-Barandela\inst{20}
\and
J.~Aumont\inst{63}
\and
C.~Baccigalupi\inst{88}
\and
A.~J.~Banday\inst{98, 10}
\and
R.~B.~Barreiro\inst{70}
% ...
}
\institute{\small
APC, AstroParticule et Cosmologie, Universit\'{e} Paris Diderot, CNRS/IN2P3, CEA/lrfu, Observatoire de Paris, Sorbonne Paris Cit\'{e}, 10, rue Alice Domon et L\'{e}onie Duquet, 75205 Paris Cedex 13, France\\
\and
Aalto University Mets\"{a}hovi Radio Observatory, Mets\"{a}hovintie 114, FIN-02540 Kylm\"{a}l\"{a}, Finland\\
\and
African Institute for Mathematical Sciences, 6-8 Melrose Road, Muizenberg, Cape Town, South Africa\\
\and
Agenzia Spaziale Italiana Science Data Center, Via del Politecnico snc, 00133, Roma, Italy\\
\and
Agenzia Spaziale Italiana, Viale Liegi 26, Roma, Italy\\
\and
Astrophysics Group, Cavendish Laboratory, University of Cambridge, J J Thomson Avenue, Cambridge CB3 0HE, U.K.\\
\and
Astrophysics \& Cosmology Research Unit, School of Mathematics, Statistics \& Computer Science, University of KwaZulu-Natal, Westville Campus, Private Bag X54001, Durban 4000, South Africa\\
\and
Atacama Large Millimeter/submillimeter Array, ALMA Santiago Central Offices, Alonso de Cordova 3107, Vitacura, Casilla 763 0355, Santiago, Chile\\
\and
CITA, University of Toronto, 60 St. George St., Toronto, ON M5S 3H8, Canada\\
\and
CNRS, IRAP, 9 Av. colonel Roche, BP 44346, F-31028 Toulouse cedex 4, France\\
\and
California Institute of Technology, Pasadena, California, U.S.A.\\
\and
Centre for Theoretical Cosmology, DAMTP, University of Cambridge, Wilberforce Road, Cambridge CB3 0WA, U.K.\\
\and
Centro de Astrof\'{\i}sica, Universidade do Porto, Rua das Estrelas, 4150-762 Porto, Portugal\\
\and
Centro de Estudios de F\'{i}sica del Cosmos de Arag\'{o}n (CEFCA), Plaza San Juan, 1, planta 2, E-44001, Teruel, Spain\\
\and
Computational Cosmology Center, Lawrence Berkeley National Laboratory, Berkeley, California, U.S.A.\\
\and
Consejo Superior de Investigaciones Cient\'{\i}ficas (CSIC), Madrid, Spain\\
\and
DSM/Irfu/SPP, CEA-Saclay, F-91191 Gif-sur-Yvette Cedex, France\\
\and
DTU Space, National Space Institute, Technical University of Denmark, Elektrovej 327, DK-2800 Kgs. Lyngby, Denmark\\
\and
D\'{e}partement de Physique Th\'{e}orique, Universit\'{e} de Gen\`{e}ve, 24, Quai E. Ansermet,1211 Gen\`{e}ve 4, Switzerland\\
\and
Departamento de F\'{\i}sica Fundamental, Facultad de Ciencias, Universidad de Salamanca, 37008 Salamanca, Spain\\
\and
Departamento de F\'{\i}sica, Universidad de Oviedo, Avda. Calvo Sotelo s/n, Oviedo, Spain\\
\and
Department of Astronomy and Astrophysics, University of Toronto, 50 Saint George Street, Toronto, Ontario, Canada\\
\and
Department of Astrophysics/IMAPP, Radboud University Nijmegen, P.O. Box 9010, 6500 GL Nijmegen, The Netherlands\\
\and
Department of Electrical Engineering and Computer Sciences, University of California, Berkeley, California, U.S.A.\\
\and
Department of Physics \& Astronomy, University of British Columbia, 6224 Agricultural Road, Vancouver, British Columbia, Canada\\
\and
Department of Physics and Astronomy, Dana and David Dornsife College of Letter, Arts and Sciences, University of Southern California, Los Angeles, CA 90089, U.S.A.\\
\and
Department of Physics and Astronomy, University College London, London WC1E 6BT, U.K.\\
\and
Department of Physics, Florida State University, Keen Physics Building, 77 Chieftan Way, Tallahassee, Florida, U.S.A.\\
\and
Department of Physics, Gustaf H\"{a}llstr\"{o}min katu 2a, University of Helsinki, Helsinki, Finland\\
\and
Department of Physics, Princeton University, Princeton, New Jersey, U.S.A.\\
\and
Department of Physics, University of California, One Shields Avenue, Davis, California, U.S.A.\\
\and
Department of Physics, University of California, Santa Barbara, California, U.S.A.\\
\and
Department of Physics, University of Illinois at Urbana-Champaign, 1110 West Green Street, Urbana, Illinois, U.S.A.\\
\and
Dipartimento di Fisica e Astronomia G. Galilei, Universit\`{a} degli Studi di Padova, via Marzolo 8, 35131 Padova, Italy\\
\and
Dipartimento di Fisica e Scienze della Terra, Universit\`{a} di Ferrara, Via Saragat 1, 44122 Ferrara, Italy\\
\and
Dipartimento di Fisica, Universit\`{a} La Sapienza, P. le A. Moro 2, Roma, Italy\\
\and
Dipartimento di Fisica, Universit\`{a} degli Studi di Milano, Via Celoria, 16, Milano, Italy\\
\and
Dipartimento di Fisica, Universit\`{a} degli Studi di Trieste, via A. Valerio 2, Trieste, Italy\\
\and
Dipartimento di Fisica, Universit\`{a} di Roma Tor Vergata, Via della Ricerca Scientifica, 1, Roma, Italy\\
\and
Discovery Center, Niels Bohr Institute, Blegdamsvej 17, Copenhagen, Denmark\\
\and
Dpto. Astrof\'{i}sica, Universidad de La Laguna (ULL), E-38206 La Laguna, Tenerife, Spain\\
\and
European Southern Observatory, ESO Vitacura, Alonso de Cordova 3107, Vitacura, Casilla 19001, Santiago, Chile\\
\and
European Space Agency, ESAC, Planck Science Office, Camino bajo del Castillo, s/n, Urbanizaci\'{o}n Villafranca del Castillo, Villanueva de la Ca\~{n}ada, Madrid, Spain\\
\and
European Space Agency, ESTEC, Keplerlaan 1, 2201 AZ Noordwijk, The Netherlands\\
\and
Finnish Centre for Astronomy with ESO (FINCA), University of Turku, V\"{a}is\"{a}l\"{a}ntie 20, FIN-21500, Piikki\"{o}, Finland\\
\and
Haverford College Astronomy Department, 370 Lancaster Avenue, Haverford, Pennsylvania, U.S.A.\\
\and
Helsinki Institute of Physics, Gustaf H\"{a}llstr\"{o}min katu 2, University of Helsinki, Helsinki, Finland\\
\and
INAF - Osservatorio Astrofisico di Catania, Via S. Sofia 78, Catania, Italy\\
\and
INAF - Osservatorio Astronomico di Padova, Vicolo dell'Osservatorio 5, Padova, Italy\\
\and
INAF - Osservatorio Astronomico di Roma, via di Frascati 33, Monte Porzio Catone, Italy\\
\and
INAF - Osservatorio Astronomico di Trieste, Via G.B. Tiepolo 11, Trieste, Italy\\
\and
INAF Istituto di Radioastronomia, Via P. Gobetti 101, 40129 Bologna, Italy\\
\and
INAF/IASF Bologna, Via Gobetti 101, Bologna, Italy\\
\and
INAF/IASF Milano, Via E. Bassini 15, Milano, Italy\\
\and
INFN, Sezione di Bologna, Via Irnerio 46, I-40126, Bologna, Italy\\
\and
INFN, Sezione di Roma 1, Universit\`{a} di Roma Sapienza, Piazzale Aldo Moro 2, 00185, Roma, Italy\\
\and
IPAG: Institut de Plan\'{e}tologie et d'Astrophysique de Grenoble, Universit\'{e} Joseph Fourier, Grenoble 1 / CNRS-INSU, UMR 5274, Grenoble, F-38041, France\\
\and
IUCAA, Post Bag 4, Ganeshkhind, Pune University Campus, Pune 411 007, India\\
\and
Imperial College London, Astrophysics group, Blackett Laboratory, Prince Consort Road, London, SW7 2AZ, U.K.\\
\and
Infrared Processing and Analysis Center, California Institute of Technology, Pasadena, CA 91125, U.S.A.\\
\and
Institut N\'{e}el, CNRS, Universit\'{e} Joseph Fourier Grenoble I, 25 rue des Martyrs, Grenoble, France\\
\and
Institut Universitaire de France, 103, bd Saint-Michel, 75005, Paris, France\\
\and
Institut d'Astrophysique Spatiale, CNRS (UMR8617) Universit\'{e} Paris-Sud 11, B\^{a}timent 121, Orsay, France\\
\and
Institut d'Astrophysique de Paris, CNRS (UMR7095), 98 bis Boulevard Arago, F-75014, Paris, France\\
\and
Institute for Space Sciences, Bucharest-Magurale, Romania\\
\and
Institute of Astronomy and Astrophysics, Academia Sinica, Taipei, Taiwan\\
\and
Institute of Astronomy, University of Cambridge, Madingley Road, Cambridge CB3 0HA, U.K.\\
\and
Institute of Theoretical Astrophysics, University of Oslo, Blindern, Oslo, Norway\\
\and
Instituto de Astrof\'{\i}sica de Canarias, C/V\'{\i}a L\'{a}ctea s/n, La Laguna, Tenerife, Spain\\
\and
Instituto de F\'{\i}sica de Cantabria (CSIC-Universidad de Cantabria), Avda. de los Castros s/n, Santander, Spain\\
\and
Jet Propulsion Laboratory, California Institute of Technology, 4800 Oak Grove Drive, Pasadena, California, U.S.A.\\
\and
Jodrell Bank Centre for Astrophysics, Alan Turing Building, School of Physics and Astronomy, The University of Manchester, Oxford Road, Manchester, M13 9PL, U.K.\\
\and
Kavli Institute for Cosmology Cambridge, Madingley Road, Cambridge, CB3 0HA, U.K.\\
\and
LAL, Universit\'{e} Paris-Sud, CNRS/IN2P3, Orsay, France\\
\and
LERMA, CNRS, Observatoire de Paris, 61 Avenue de l'Observatoire, Paris, France\\
\and
Laboratoire AIM, IRFU/Service d'Astrophysique - CEA/DSM - CNRS - Universit\'{e} Paris Diderot, B\^{a}t. 709, CEA-Saclay, F-91191 Gif-sur-Yvette Cedex, France\\
\and
Laboratoire Traitement et Communication de l'Information, CNRS (UMR 5141) and T\'{e}l\'{e}com ParisTech, 46 rue Barrault F-75634 Paris Cedex 13, France\\
\and
Laboratoire de Physique Subatomique et de Cosmologie, Universit\'{e} Joseph Fourier Grenoble I, CNRS/IN2P3, Institut National Polytechnique de Grenoble, 53 rue des Martyrs, 38026 Grenoble cedex, France\\
\and
Laboratoire de Physique Th\'{e}orique, Universit\'{e} Paris-Sud 11 \& CNRS, B\^{a}timent 210, 91405 Orsay, France\\
\and
Lawrence Berkeley National Laboratory, Berkeley, California, U.S.A.\\
\and
Max-Planck-Institut f\"{u}r Astrophysik, Karl-Schwarzschild-Str. 1, 85741 Garching, Germany\\
\and
McGill Physics, Ernest Rutherford Physics Building, McGill University, 3600 rue University, Montr\'{e}al, QC, H3A 2T8, Canada\\
\and
MilliLab, VTT Technical Research Centre of Finland, Tietotie 3, Espoo, Finland\\
\and
Niels Bohr Institute, Blegdamsvej 17, Copenhagen, Denmark\\
\and
Observational Cosmology, Mail Stop 367-17, California Institute of Technology, Pasadena, CA, 91125, U.S.A.\\
\and
Optical Science Laboratory, University College London, Gower Street, London, U.K.\\
\and
SB-ITP-LPPC, EPFL, CH-1015, Lausanne, Switzerland\\
\and
SISSA, Astrophysics Sector, via Bonomea 265, 34136, Trieste, Italy\\
\and
School of Physics and Astronomy, Cardiff University, Queens Buildings, The Parade, Cardiff, CF24 3AA, U.K.\\
\and
School of Physics and Astronomy, University of Nottingham, Nottingham NG7 2RD, U.K.\\
\and
Space Research Institute (IKI), Russian Academy of Sciences, Profsoyuznaya Str, 84/32, Moscow, 117997, Russia\\
\and
Space Sciences Laboratory, University of California, Berkeley, California, U.S.A.\\
\and
Special Astrophysical Observatory, Russian Academy of Sciences, Nizhnij Arkhyz, Zelenchukskiy region, Karachai-Cherkessian Republic, 369167, Russia\\
\and
Stanford University, Dept of Physics, Varian Physics Bldg, 382 Via Pueblo Mall, Stanford, California, U.S.A.\\
\and
Sub-Department of Astrophysics, University of Oxford, Keble Road, Oxford OX1 3RH, U.K.\\
\and
Theory Division, PH-TH, CERN, CH-1211, Geneva 23, Switzerland\\
\and
UPMC Univ Paris 06, UMR7095, 98 bis Boulevard Arago, F-75014, Paris, France\\
\and
Universit\'{e} de Toulouse, UPS-OMP, IRAP, F-31028 Toulouse cedex 4, France\\
% ...
}



 \abstract {
 We have constructed the first all-sky map of the thermal Sunyaev-Zeldovich
 (tSZ) effect by applying specifically tailored component separation
 algorithms to the 100 to 857\,GHz frequency channel maps from the \Planck\
 survey.  This map shows an obvious galaxy cluster tSZ signal that is well
 matched with blindly detected clusters in the \Planck\ SZ catalogue.
 To characterize the signal in the tSZ map we have computed its angular
 power spectrum.
 At large angular scales ($\ell < 60$), the major foreground contaminant is
 the diffuse thermal dust emission.  At small angular scales ($\ell > 500$)
 the clustered cosmic infrared background (CIB) and residual point sources
 are the major contaminants.
 These foregrounds are carefully modelled and subtracted. 
 We thus measure the tSZ power spectrum over angular scales
 $0.17^{\circ} \la \theta \la 3.0^{\circ}$ that were previously
 unexplored.
 The measured tSZ power spectrum is consistent with that expected from the
 \Planck\ catalogue of SZ sources, with clear evidence of additional signal
 from unresolved clusters and, potentially, diffuse warm baryons.
 Marginalized band-powers of the \Planck\ tSZ power spectrum and the
 best-fit model are given.
 We use the tSZ power spectrum to obtain the following cosmological
 constraint:
 $\sigma_8(\Omega_{\mathrm{m}}/0.28)^{0.40}=0.784 \pm 0.016
 \left( 68 \ \mathrm{ \% \ C.L.} \right)$.
 This result is in weak tension (at 2.7$\,\sigma$) with the 
 \Planck\ measurements of the primary anisotropies in the cosmic microwave
 background.
 The non-Gaussianity of the Compton parameter map is further characterized
 by computing its 1D probability distribution function and its bispectrum.
 These are used to place additional independent constraints on $\sigma_{8}$. 
  }


\authorrunning{Planck Collaboration}
\titlerunning{Cosmology from the \Planck\ all-sky Compton $y$ parameter}

\maketitle
\clearpage
\section{Introduction}
\label{sec:introduction}

This paper, one of a set associated with the 2013 release of data from the
\Planck\footnote{\Planck\ (\url{http://www.esa.int/Planck}) is a project of
the European Space Agency (ESA) with instruments provided by two scientific
consortia funded by ESA member states (in particular the lead countries France
and Italy), with contributions from NASA (USA) and telescope reflectors
provided by a collaboration between ESA and a scientific consortium led and
funded by Denmark.} mission \citep{planck2013-p01}, describes the
construction of a Compton $y$ parameter map and the determination of
its angular power spectrum and high-order statistics.

The thermal Sunyaev-Zeldovich (tSZ) effect \citep{SZ}, produced by the
inverse Compton scattering of cosmic microwave background (CMB)
photons by hot electrons along the
line of sight, has proved to be a major tool for studying the
physics of clusters of galaxies as well as structure formation in the
Universe.  In particular, tSZ-selected catalogues of clusters of
galaxies have been provided by various experiments including the
\Planck\ satellite \citep{planck2011-5.1a,planck2013-p05a}, the
Atacama Cosmology Telescope
\citep[ACT,][]{Hasselfield:2013p2303} and the
South Pole Telescope \citep[SPT,][]{Reichardt:2013p2252}.  These
catalogues and their associated sky surveys have been used to study
the physics of clusters of galaxies
\citep{planck2011-5.2c,planck2011-5.2b,planck2011-5.2a} and their
cosmological implications
\citep{planck2013-p15,Benson:2013p2263,Das:2013p2167,Wilson:2012p2102,Mak:2012p2066}.

The study of number counts and their evolution with
redshift using tSZ detected clusters of galaxies is
an important cosmological test
\citep{car02,Dunkley:2013p2181,Benson:2013p2263,planck2013-p15}.
The measurement of the tSZ effect power spectrum has been
proposed by \citet{Komatsu:2002p1799} as a complement to the counts.
One advantage of using the tSZ
angular power spectrum over cluster counts is that no explicit
measurement of cluster masses is required.   
Also, lower mass, and
therefore fainter, clusters, which may not be detected as
individual objects, contribute to this statistical signal
\citep{Battaglia2010,Shaw2010}.
However, significant drawbacks of using the tSZ
angular power spectrum include potential contamination from point
sources \citep{RubinoMartin:2003p1790,Taburet2010a} and other
foregrounds. 

To date, measurements of the tSZ power
spectrum are only available from high resolution CMB-oriented experiments
like ACT \citep{Sievers:2013p2161} and SPT
\citep{2012ApJ...755...70R}. In these studies, constraints on the amplitude of
the tSZ power spectrum at $\ell=3000$ are obtained by fitting a tSZ template
in addition to other components (i.e., CMB, radio and infrared point-source
and clustered cosmic infrared background, CIB) to the measured
total power spectrum. These constraints are obtained at angular
scales where the tSZ signal dominates over the CMB, but at these same
scales the contamination from point sources and the clustered CIB is
important and may affect the measured tSZ signal. Moreover, the
scales probed are particularly sensitive to the uncertainties in
modelling the intracluster medium (ICM) over a broad range of masses
and redshifts, and at large cluster-centric radii
\citep{Battaglia2010}. Recent work, using hydrodynamical simulations
\citep{Battaglia2010,Battaglia:2012p1841} $N$-body
simulations plus semi-analytic gas models \citep{Trac:2011p1795} and
purely analytic models \citep{Shaw2010}, have significantly reduced the
tension between the observed and predicted values. However the
distribution of amplitudes between different models and simulations is
still significantly larger than the measurement errors, degrading the
constraints that can be placed on cosmological parameters with these methods \citep{Dunkley:2013p2181,Reichardt:2013p2252}.

In addition to the power spectrum, and
\citep[as pointed out in][]{RubinoMartin:2003p1790},
the skewness or, equivalently, the
bispectrum of the tSZ signal is a powerful and independent tool to
study and to isolate the signal of clusters, separating it from the
contribution of radio and IR sources. Recently,
\citet{Bhattacharya:2012p2458} showed that the bispectrum of the tSZ
effect signal is dominated by massive clusters at intermediate
redshifts, for which high-precision X-ray observations exist. This
contrasts with the power spectrum, where the signal mainly comes from
the lower mass and higher redshift groups and clusters
\citep[e.g.,][]{Trac:2011p1795}. The theoretical uncertainty in the
tSZ bispectrum is thus expected to be significantly smaller than that
of the SZ power spectrum. Combined measurements of the power spectrum
and the bispectrum can thus be used to distinguish the contribution to
the power spectrum from different cluster masses and redshift
ranges. The bispectrum amplitude scales as $\sigma_{8}^{10-12}$
\citep{Bhattacharya:2012p2458}.  
Measurements of the tSZ bispectrum have been reported by the SPT
collaboration \citep{2013arXiv1303.3535C}. Alternatively,
\cite{Wilson:2012p2102} used the unnormalized skewness of the tSZ
fluctuations, $\langle T^{3}(\mathbf{n})\rangle$,
which scales approximately as $\sigma_{8}^{11}$,
to obtain an independent determination of $\sigma_8$.

Thanks to its all-sky coverage and unprecedented wide
frequency range, \Planck\ has the unique ability to
produce an all-sky tSZ Compton parameter ($y$\/) map and an accurate
measurement of the tSZ power spectrum at intermediate and large
angular scales, for which the tSZ fluctuations are almost insensitive to
the cluster core physics.  The \Planck\ Compton parameter map
also offers the possibility of studying the properties
of the non-Gaussianity of the tSZ signal using higher order statistical
estimators, such as the skewness and the bispectrum. In this paper we construct
a tSZ all-sky map from the individual \Planck\ frequency maps and
compute its power spectrum, its 1D probability density
function (1D PDF), and the associated bispectrum. 
 
The paper is structured as
follows. Section~\ref{datasimu} describes the \Planck\ data used to
compute the tSZ all-sky map and the simulations used to characterize
it.  We discuss details of the modelling of the tSZ effect power
spectrum and bispectrum in Sect.~\ref{sec:theory}.  In
Sect.~\ref{sec:allskymap} we present the \Planck\ all-sky Compton
parameter map. Section~\ref{sec:powerspec} describes the power
spectrum analysis.  Cross-checks using high-order statistics are
presented in Sect.~\ref{sec:higorderstat}. Cosmological
interpretation of the results is discussed in
Sect.~\ref{sec:cosmo}, and we present our conclusions in
Sect.~\ref{conclusions}.

\section{Data and  simulations}
\label{datasimu}

\begin{table}[tmb]
\begingroup
\newdimen\tblskip \tblskip=5pt
\caption{\label{table:summary} Conversion factors for tSZ Compton parameter
$y$ to CMB temperature units and the FWHM of the beam of the
\Planck\ channel maps.}                          
\nointerlineskip
\vskip -4mm
\footnotesize
\setbox\tablebox=\vbox{
   \newdimen\digitwidth 
   \setbox0=\hbox{\rm 0} 
   \digitwidth=\wd0 
   \catcode`*=\active 
   \def*{\kern\digitwidth}
   \newdimen\dpwidth 
   \setbox0=\hbox{.} 
   \dpwidth=\wd0 
   \catcode`!=\active 
   \def!{\kern\dpwidth}
\halign{\hbox to 1.5cm{#\leaderfil}\tabskip 1em&
     \hfil#\hfil \tabskip 1em&
     \hfil#\hfil \tabskip 0em \cr
\noalign{\doubleline}
\omit  Frequency &  $ T_{\mathrm{CMB}}  \ g (\nu)$ & FWHM\cr
\omit\hfil$[$GHz$]$\hfil& $[$K$_{\mathrm{CMB}}]$ & [arcmin]\cr
\noalign{\vskip 3pt\hrule\vskip 5pt}
100& $-4.031$& 9.66\cr
143& $-2.785$& 7.27\cr
217& $*0.187$& 5.01\cr
353& $*6.205$& 4.86\cr
545& $14.455$& 4.84\cr
857& $26.335$& 4.63\cr
\noalign{\vskip 3pt\hrule\vskip 5pt}
}
}
\endPlancktable 
\endgroup
\end{table}

\subsection{The \Planck\ data}
\label{subsec:planckdata}

This paper is based on the first 15.5 months of \Planck's mission,
corresponding to more than two full-sky surveys.  We refer to
\citet{planck2013-p02}, \citet{planck2013-p02a}, \citet{planck2013-p02b},
\citet{planck2013-p03f}, \citet{planck2013-p03d},
and \citet{planck2013-p03} for the generic scheme of time-ordered
information processing and map-making, as well as for
the technical characteristics of the \Planck\ frequency maps.  The
\Planck\ channel maps are provided in {\tt HEALPix} \citep{healpix}
pixelization scheme at $N_{\mathrm{side}}=2048$. An error map is
associated with each channel map and is obtained from the difference of maps made from the first and second half of each ring (stable pointing period).
 The difference maps, called
half-ring or null maps, are mainly free from astrophysical emission and they
are a good representation of the statistical instrumental noise. Null
maps have also been used to estimate the noise in the final Compton
parameter maps.  Here we approximate the \Planck\ effective beams by 
circular Gaussians \citep{planck2013-p02d,planck2013-p03c} The FWHM values
for each frequency channel are given in Table~\ref{table:summary}.
Although tests have been performed using both LFI and HFI channel
maps, the work presented here is based mostly on HFI data.

\subsection{{\tt FFP6} Simulations}

We also use simulated \Planck\ frequency maps
obtained from the  Full Focal Plane ({\tt FFP6})
simulations, which are described in the \Planck\ Explanatory
Supplement \citep{planck2013-p28}.
These simulations include the most relevant sky components at microwave and
millimetre frequencies, based on foregrounds from the \Planck\ Sky Model
\citep[PSM,][]{Delabrouille2012}:
CMB; thermal SZ effect; diffuse Galactic
emissions (synchrotron, free-free, thermal and spinning dust and CO); radio
and infrared point sources, and the clustered CIB.
The simulated tSZ signal was
constructed using hydrodynamical simulations of clusters of galaxies
up to redshift 0.3, completed with pressure profile-based simulations
of individual clusters of galaxies randomly drawn on the sky.  The
noise in the maps was obtained from realizations of Gaussian random 
noise in the time domain and therefore accounts for noise
inhomogeneities in the maps.

\section{Modelling the tSZ effect}
\label{sec:theory}

The thermal SZ Compton parameter in a given direction, $\vec{n}$, is 
\begin{equation}
y (\vec{n}) = \int n_{\mathrm{e}} \frac{k_{\mathrm{B}}
 T_{\mathrm{e}}}{m_{\mathrm{e}} c^{2} } \sigma_{\mathrm{T}} \  \mathrm{d}s,
\end{equation}
where $k_{\mathrm{B}}$ is the Boltzmann contant, $m_{\mathrm{e}}$ the electron
mass, $\sigma_{\mathrm{T}}$ the Thomson cross-section,
d$s$ the distance along the line of sight, $\vec{n}$, and
$n_{\mathrm{e}}$ and $T_{\mathrm{e}}$ are the electron number density and
temperature.

In units of CMB temperature the contribution of the tSZ effect
to the \Planck\ maps for a given frequency $\nu$ is
\begin{equation}
\frac{\Delta T}{T_{\mathrm{CMB}} }= g(\nu) \ y.
\end{equation}
Neglecting relativistic corrections we have
$g(\nu) = [x \  \coth(x/2) - 4 ]$,
with $ x=h \nu/(k_{\mathrm{B}} T_{\mathrm{CMB}})$.  Table~\ref{table:summary}
shows the conversion factors for Compton parameter to CMB temperature,
K$_{\mathrm{CMB}}$, for each frequency channel after integrating over
the bandpass.

\subsection{tSZ power spectrum}
\label{poowerspecth}
Decomposing the map in spherical harmonics, $Y_{\ell m}$, we obtain
\begin{equation}
y(\vec{n}) = \sum_{\ell m} \ y_{\ell m} \ Y_{\ell m} (\vec{n}).
\end{equation}
Thus, the angular power spectrum of the Compton parameter map is
\begin{equation}
C^{\mathrm{tSZ}}_{\ell} = \frac{1}{2\ \ell +1}
 \sum_{m} y_{\ell m}  y^{*}_{\ell m}.
\end{equation}
Note that $C^{\mathrm{tSZ}}_{\ell}$ is a dimensionless quantity here,
like $y$.

To model the tSZ power spectrum we consider a 2-halo model to
account for intra-halo and inter-halo correlations:
\begin{equation}
C_\ell^{\mathrm{SZ}} = C_\ell^{\mathrm{1halo}}+ C_\ell^{\mathrm{2halos}}.
\end{equation}
The 1-halo term, also known as the Poissonian contribution, can be
computed by summing the square of the Fourier transform of the
projected SZ profile, weighted by the number density of clusters of a
given mass and redshift \citep{Komatsu:2002p1799}:
\begin{equation}
C_\ell^{\mathrm{1halo}}
 = \int_0^{z_\mathrm{max}}dz\frac{dV_\mathrm{c}}
 {dzd\Omega}\int_{M_{\mathrm{min}}}^{M_{\mathrm{max}}}dM
 \frac{dn(M,z)}{dM}\left|\tilde{y_\ell}(M,z)\right|^2,
\end{equation}
where $dV_{\mathrm{c}}/(dz  d\Omega)$ is the comoving volume per
unit redshift and solid angle and
$n(M,z)dM \ dV_{\mathrm{c}}/(dz d\Omega)$ is the probability of having
a galaxy cluster of mass $M$ at a redshift $z$ in the direction
$d\Omega$. The quantity
$\tilde{y}_\ell=\tilde{y}_\ell(M,z)$ is the 2D Fourier
transform on the sphere of the 3D radial profile of the Compton
$y$-parameter of individual clusters,
\begin{equation}
\tilde{y}_\ell(M,z) = \frac{4 \pi r_{\mathrm{s}}}{l_{\mathrm{s}}^2} \left( \frac{\sigma_{\mathrm{T}}}{m_{\mathrm{e}}c^{2}}\right) \int_{0}^{\infty} \ dx \ x^{2} P_{\mathrm{e}} (M,z,x) \frac{\sin(\ell_{x}/\ell_{\mathrm{s}})}{\ell_{x}/\ell_{\mathrm{s}}} 
\end{equation}
where $x=r/r_{\mathrm{s}}$, $\ell_{\mathrm{s}} =
D_{\mathrm{A}}(z)/r_{\mathrm{s}}$, $r_{\mathrm{s}}$ is the scale
radius of the 3D pressure profile, $D_{\mathrm{A}}(z)$ is the angular diameter
distance to redshift $z$ and $P_{\mathrm{e}}$ is the electron pressure
profile.
 
The 2-halo term is obtained by computing the correlation between two
different halos \citep{Komatsu:1999p2519,Diego2004,Taburet11}:
\begin{eqnarray}
\nonumber
C_\ell^{\mathrm{2halos}}&=&\int_0^{z_\mathrm{max}}dz
 \frac{dV_\mathrm{c}}{dzd\Omega}  \times \\
 & & \left[ \int_{M_{\mathrm{min}}}^{M_{\mathrm{max}}}dM
 \frac{dn(M,z)}{dM}\left|\tilde{y}_\ell(M,z)\right| \ B(M,z) \right]^2
 \ P(k,z),
\label{twohalomodel}
\end{eqnarray}
where $P(k,z)$ is the 3D matter power spectrum at redshift $z$.  Here
$B(M,z)$ is the time-dependent linear bias factor that relates the
matter power spectrum, $P(k,z)$, to the power spectrum of the cluster
correlation function. Following
\citet[][see also \citealt{Mo:1996p2742}]{Komatsu:1999p2519} we adopt
$B(M,z) = 1 + (\nu^{2}(M,z)-1)/\delta_{\mathrm{c}}(z) $, where $\nu(M,z) =
\delta_{\mathrm{c}}(M)/D(z)\sigma(M)$, $\sigma(M)$ is the present-day rms mass
fluctuation, $D(z)$ is the linear growth factor, and
$\delta_{\mathrm{c}}(z)$ is the threshold over-density of spherical
collapse.

Finally, we compute the tSZ power spectrum using the \cite{Tinker2008}
mass function $dn(M,z)/dM$ including an observed-to-true mass bias of
20\%, as discussed in detail in \citet{planck2013-p15}, and we model the
SZ Compton parameter using the pressure profile of \cite{Arnaud2010}.
This approach is adopted in order to be consistent with the ingredients
of the cluster number count analysis in \citet{planck2013-p15}.

\subsection{$N\mathrm{th}$ moment of the tSZ field}
\label{nthmoment}
To calculate the $N\mathrm{th}$ moment of the tSZ field, we assume,
to first order, that the distribution of clusters on the sky can be
adequately described by a Poisson distribution corresponding to the
1-halo term. We neglect the contribution due to clustering between clusters
and their overlap \citep{Komatsu:1999p2519}. The $N\mathrm{th}$ moment is
then given
by \citep{Wilson:2012p2102}
\begin{equation}
 \int_0^{z_\mathrm{max}}dz\frac{dV_\mathrm{c}}{dzd\Omega}
 \int_{M_{\mathrm{min}}}^{M_{\mathrm{max}}}dM\frac{dn(M,z)}{dM}
 \int d^{2}\mathbf{\theta} \ {y(\mathbf{\theta},M,z)}^{N},
\end{equation}
where $y(\mathbf{\theta},M,z)$ is the integrated Compton parameter
along the line of sight for a cluster of mass $M$ at redshift $z$.

\subsection{Bispectrum}
\label{bispectrumth}

The angular bispectrum, analogous to the 3-point correlation function
in harmonic space, is the lowest-order indicator of the
non-Gaussianity of a field. It is given by
\begin{equation}
B^{m_{1} m_{2} m_{3}}_{\ell_{1} \ell_{2} \ell_{3}}
 = \left< y_{\ell_{1} m_{1}}  y_{\ell_{2} m_{2}}  y_{\ell_{3} m_{3}} \right>,
\end{equation}
where the angle-averaged quantity in the full-sky limit can be written as
\begin{equation}
b(\ell_{1},\ell_{2},\ell_{3}) = \sum_{m_{1}m_{2}m_{3}}
 \left( \begin{array}{ccc}  \ell_{1} & \ell_{2} & \ell_{3} \\
 m_{1} & m_{2} & m_{3} \end{array} \right)
 B^{m_{1} m_{2} m_{3}}_{\ell_{1} \ell_{2} \ell_{3}},
\end{equation}
which has to satisfy the conditions $m_{1}+m_{2}+m_{3}=0$,
 $\ell_{1}+\ell_{2}+\ell_{3}=\mathrm{even}$, and
 $\left|\ell_{i}-\ell_{j}\right|\leq\ell_{k}\leq\ell_{i}+\ell_{j}$,
for the Wigner $3j$ function in brackets. For illustration we
compute the bispectrum assuming a Poissonian distribution, given by 
\citep{Bhattacharya:2012p2458}
\begin{eqnarray}
\nonumber
b(\ell_{1},\ell_{2},\ell_{3}) \approx
 \sqrt{\frac{(2\ell_{1}+1)(2\ell_{2}+1)(2\ell_{3}+1)}{4 \pi}}
 \left( \begin{array}{ccc}  \ell_{1} & \ell_{2} & \ell_{3} \\
 0 & 0 & 0 \end{array} \right) \\
 \int_0^{z_\mathrm{max}}\!\!dz\frac{dV_\mathrm{c}}{dzd\Omega}
 \int_{M_{\mathrm{min}}}^{M_{\mathrm{max}}}\!\!dM\frac{dn(M,z)}{dM}
 \tilde{y}_{\ell_{1}}(M,z) \tilde{y}_{\ell_{2}}(M,z) \tilde{y}_{\ell_{3}}(M,z).
\end{eqnarray}

\section{The reconstructed all-sky tSZ map}
\label{sec:allskymap}

\subsection{Reconstruction methods}
\label{subsec:compsep}
The contribution of the tSZ effect in the \Planck\ frequency maps is
subdominant with respect to the CMB and other foreground
emissions. Furthermore, the tSZ effect from galaxy clusters is spatially
localized and leads to a highly non-Gaussian signal with respect to
that from the CMB. CMB-oriented component-separation methods
\citep{planck2013-p06} are not optimized to recover the tSZ signal. We
therefore need to use specifically tailored component separation
algorithms that are able to reconstruct the tSZ signal from the
\Planck\ frequency channel maps. These optimized all-sky component separation
techniques rely on the spatial localization of the different
astrophysical components and on their spectral diversity to separate
them. We present in the following, the results of two algorithms, {\tt
MILCA} \citep[Modified Internal Linear Combination Algorithm,][]{Hurier2010} and
{\tt NILC} \citep[Needlet Iindependent Linear
Combination,][]{Remazeilles2011}. Both are based on the well known Internal
Linear Combination (ILC) approach that searches for the linear combination
of the input maps that minimizes the variance of the final
reconstructed map under the constraint of offering unit gain to the
component of interest (here the tSZ effect, whose frequency
dependence is known). Both algorithms have been extensively tested on
simulated \Planck\ data. 

\subsubsection{{\tt MILCA}}

{\tt MILCA}~\citep{Hurier2010} uses two constraints: preservation of
the tSZ signal, assuming the tSZ spectral signature; and
removal of the CMB contamination in the final SZ map, making use
of the well known spectrum of the CMB. In addition, to compute the
weights of the linear combination, we have used the extra degrees of
freedom in the linear system to minimize residuals from other components
(two degrees of freedom) and from the noise (two additional degrees).
The noise covariance matrix was estimated from the null maps
described in Section~\ref{subsec:planckdata}. To improve the
efficiency of the {\tt MILCA} algorithm, weights are allowed to vary
as a function of multipole $\ell$, and are computed independently on
different sky regions. We have used 11 filters in $\ell$
space, with an overall transmission of one, except for $\ell < 8$. For
these large angular scales we have used a Gaussian filter to reduce
foreground contamination.  The size of the independent sky regions was
adapted to the multipole range to ensure sufficient spatial
localization at the required resolution. We used a minimum of 12
regions at low resolution and a maximum of 3072 regions at high
resolution.

\subsubsection{{\tt NILC}}

In the multi-component extensions of
{\tt NILC}~\citep{Delabrouille:2009p2535,Remazeilles2011},
initially developed to extract the CMB, the weights for component separation
(i.e., covariances) are computed independently in domains of a needlet
decomposition (in the spherical wavelet frame). The needlet decomposition
provides localization of the ILC filters both in pixel and in multipole space,
allowing us to deal with local contamination conditions varying both in position
and in scale. We imposed constraints to remove the CMB contamination and
preserve the tSZ effect.  To avoid strong foreground effects, the
Galactic plane was masked before applying {\tt NILC} to the \Planck\ frequency
maps. 
  
In both methods, we mask the brightest regions in the \Planck\ 857\,GHz
channel map, corresponding to about 33\% of the sky. We use the HFI channel
maps from 100 to 857\,GHz that are convolved to a common resolution of
10\arcm. The 857\,GHz map is mainly exploited in the internal linear
combination as a template to remove the thermal dust emission on large angular
scales. However, this induces significant CIB residuals in the tSZ map on small
scales. To avoid this contamination, while enabling efficient removal of the
diffuse thermal dust emission at large angular scales, we use the 857\,GHz
channel only for $\ell < 300$.

\subsection{Reconstructed Compton parameter $y$ map}
\label{subsec:ymaps}
\begin{figure*}
\begin{center}
\includegraphics[width=2\columnwidth,angle=180]{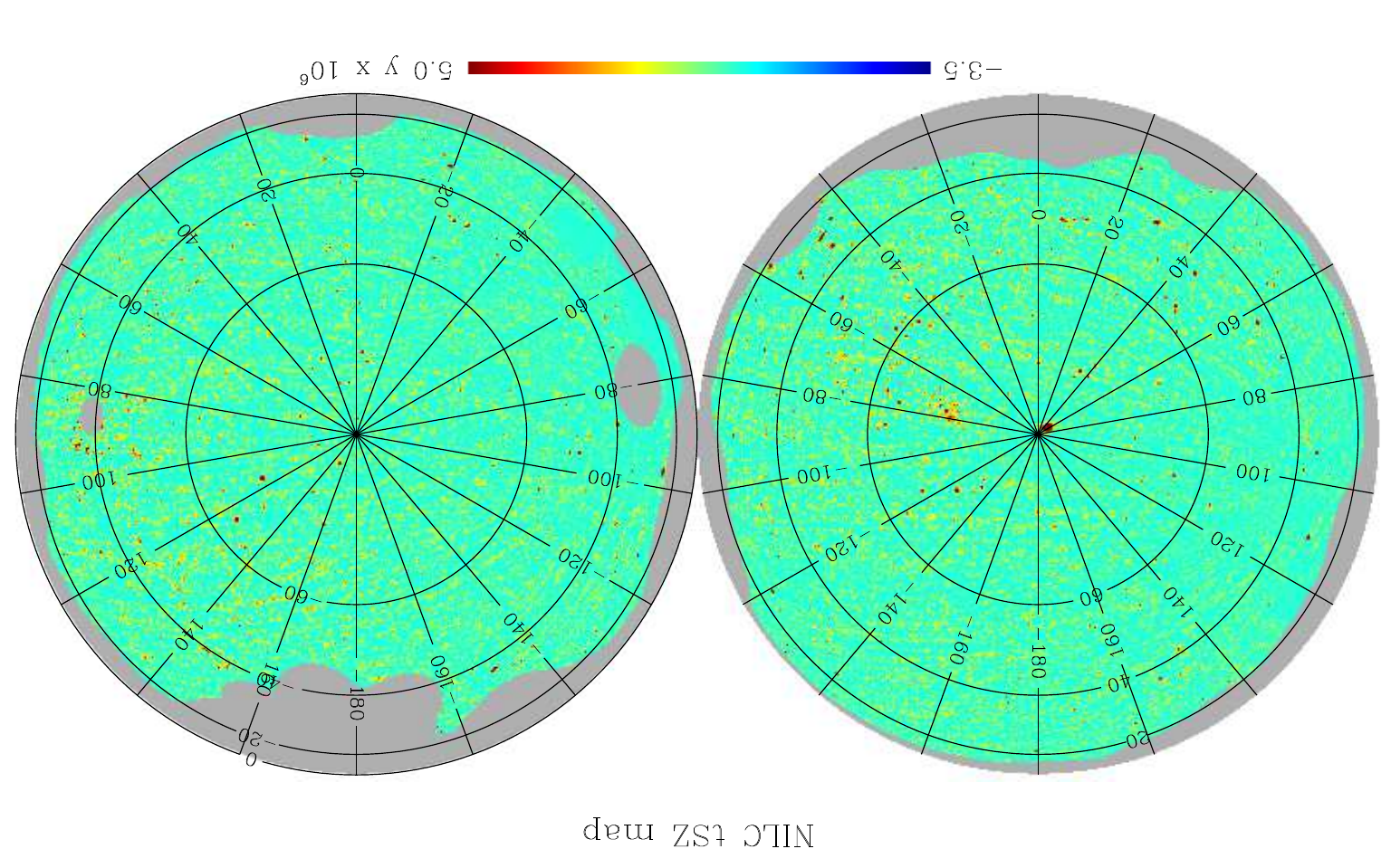}
\includegraphics[width=2\columnwidth,angle=180]{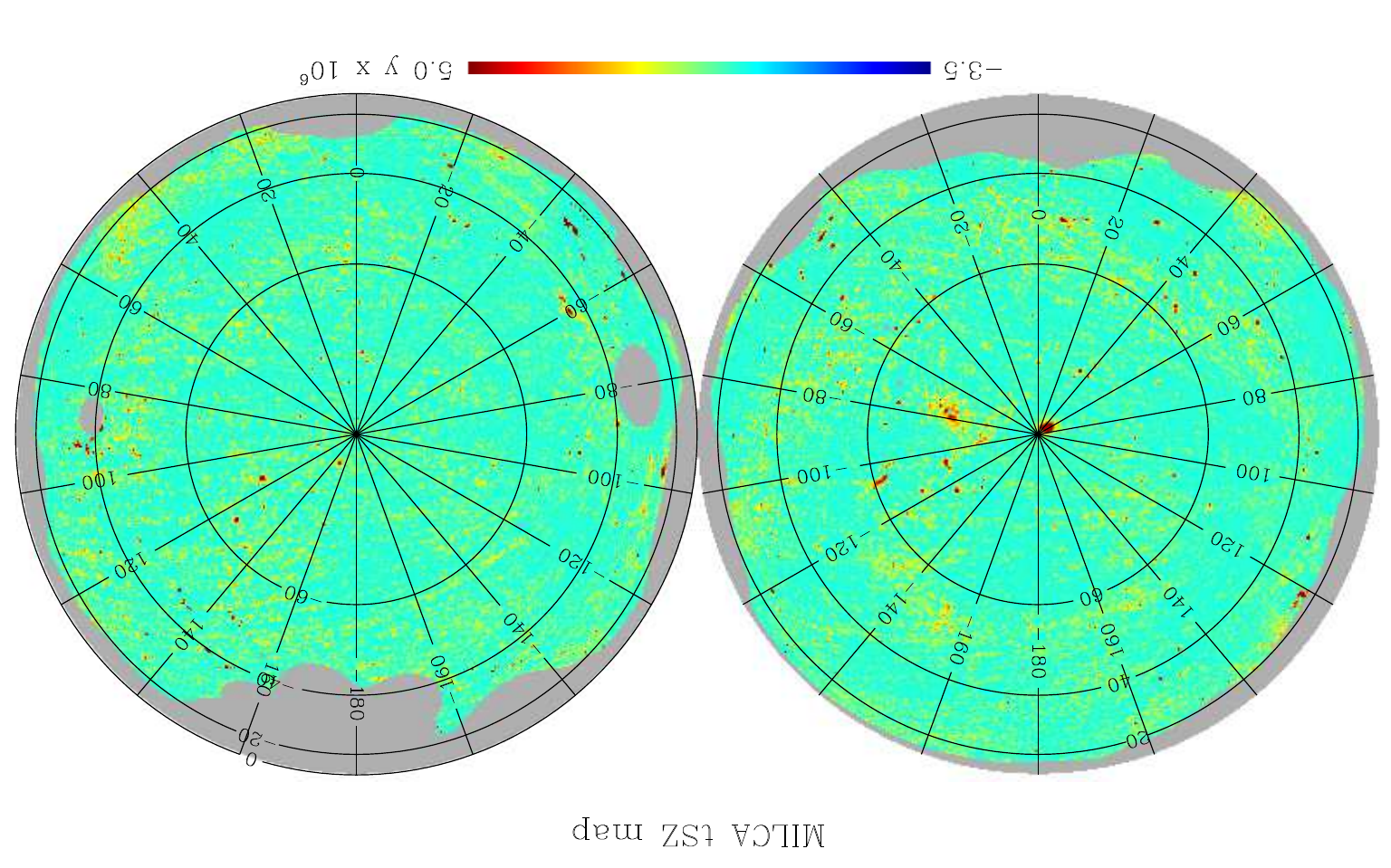}
\end{center}
\caption{Reconstructed \Planck\ all-sky Compton parameter maps for {\tt NILC}
(\textit{top}) and {\tt MILCA} (\textit{bottom}) in orthographic projections.
The apparent difference in contrast observed between the {\tt NILC} and
{\tt MILCA} maps comes from differences in the instrumental noise
contribution and foreground contamination and from the differences in the
filtering applied for display purpose to the original Compton parameter maps.
\label{fig:planck_y_map}}
\end{figure*}

\begin{figure*}
\begin{center}
\includegraphics[width=0.8\columnwidth]{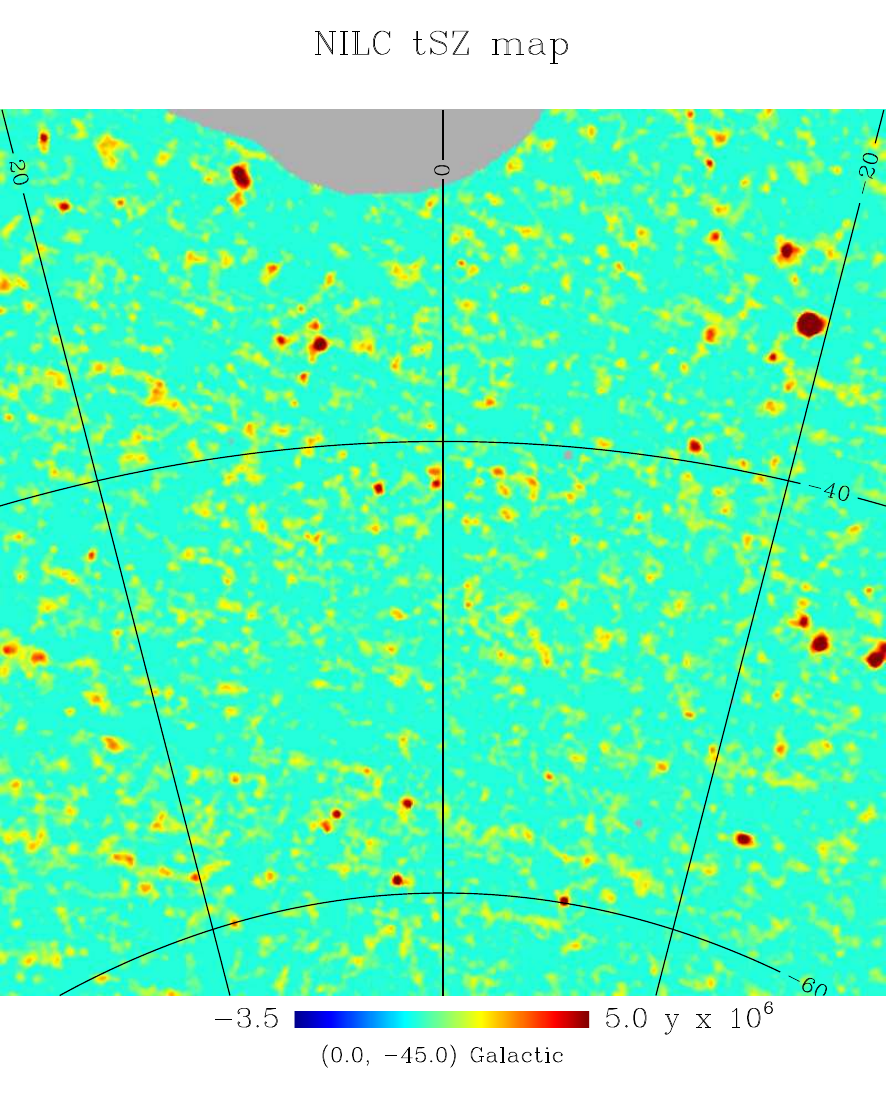}
\includegraphics[width=0.8\columnwidth]{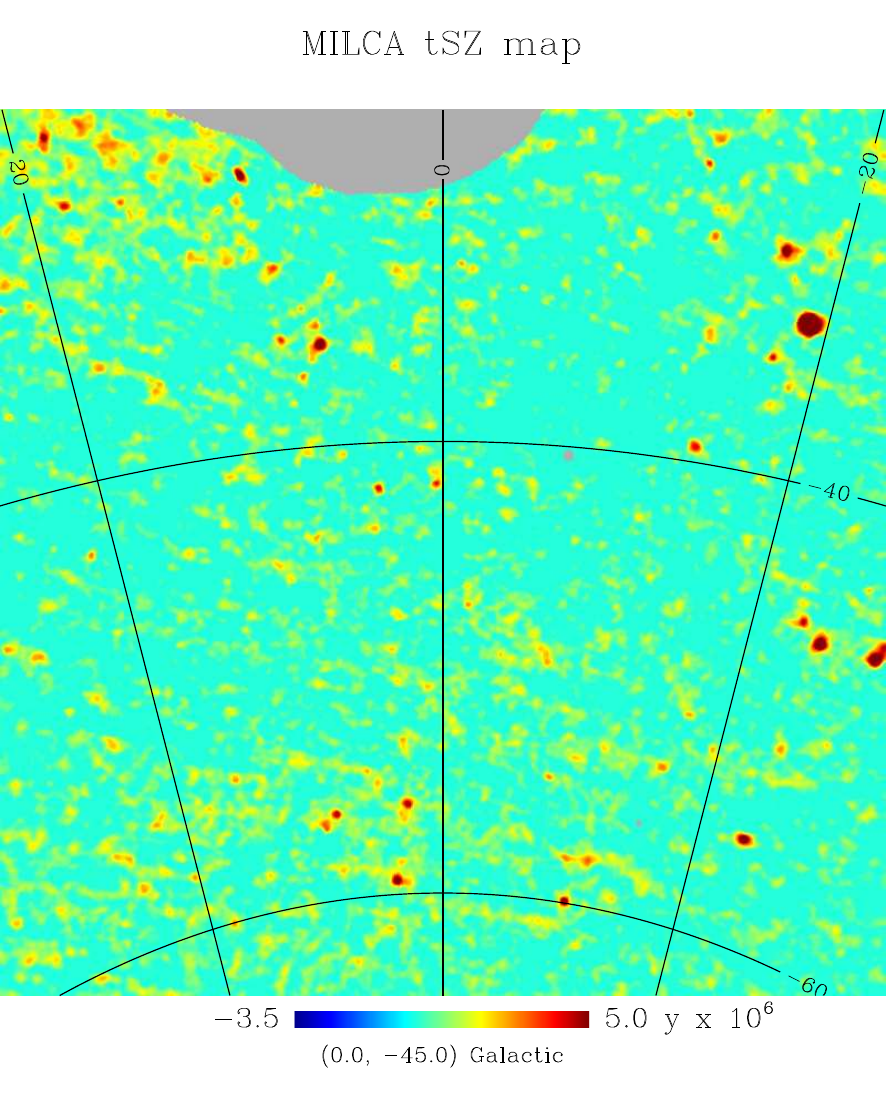}
\end{center}
\caption{A small region of the reconstructed \Planck\ all-sky Compton parameter maps for {\tt NILC} (left) and {\tt MILCA} (right) 
at intermediate Galactic latitudes in the southern sky.
\label{fig:planck_y_map_zoom}}
\end{figure*}

Figure~\ref{fig:planck_y_map} shows the reconstructed
\Planck\ all-sky Compton parameter map for {\tt NILC} (top panel) and
{\tt MILCA} (bottom panel).  For display purposes, the
maps are filtered using the procedure described in
Sect.~\ref{sec:higorderstat}. Clusters appear as positive sources: the Coma
cluster and Virgo supercluster are clearly visible near the north Galactic
pole. As mentioned above, the Galactic plane is masked in both maps,
leaving 67\% of the sky.  
Other weaker and more compact clusters are visible in the zoomed region of the
Southern cap, shown in the bottom panel of Fig.~\ref{fig:planck_y_map_zoom}.
Strong Galactic and extragalactic radio sources show up as negative bright
spots on the maps and were masked prior to any scientific analysis, as
discussed below in Sect.~\ref{subsec:pointsourcecont}. Residual Galactic
contamination is also visible around the edges of the masked area; extra
masking was performed to avoid this highly contaminated area.  The apparent
difference of contrast observed between the {\tt NILC} and {\tt MILCA}
maps comes from differences in the instrumental noise and
foreground  contamination
(the {\tt NILC} map is slightly noisier but less affected by
residual foreground emission than the {\tt MILCA} map, as discussed in
Sect.~\ref{subsec:forecont}) and from the differences in the filtering
applied for display purposes to the original Compton parameter maps,
as discussed in Sect.~\ref{methods:1D PDF}.

In addition to the full Compton parameter maps, we also produce the
so-called ``FIRST'' and ``LAST'' Compton parameter maps from the first and
second halves of the survey rings (i.e., pointing periods). These maps are
used for the power spectrum analysis 
in~Sect.~\ref{sec:powerspec}.

\subsection{Point source contamination and masking}
\label{subsec:pointsourcecont}

Point source contamination is an important issue for the cosmological
interpretation of the \Planck\ Compton parameter map. Radio sources will show up in the reconstructed tSZ 
maps as negative peaks, while infrared sources will show up as positive
peaks, mimicking the cluster signal. To avoid contamination from these 
sources we introduce a point source mask (PSMASK, hereafter). This mask is the union of the
  individual frequency point-source masks discussed
  in~\citet{planck2013-p05}. To test the reliability of this mask we
have performed a search for negative sources in the Compton parameter
maps using the {\tt MHW2} algorithm \citep{LopezCaniego:2006p2546}. We
found that all detected radio sources in the Compton parameter
maps are masked by the PSMASK. For infrared sources, estimating the
efficiency of the masking is hampered by the tSZ signal itself. The
residual contamination from point sources is discussed in
Sects.~\ref{subsec:forecont} and~\ref{sec:higorderstat}. It is
also important to note that the PSMASK may also exclude some
clusters of galaxies. This is particularly true in the case of clusters
with strong central radio sources, such as the Perseus
cluster~\citep[see][]{planck2013-p05a}.

\subsection{tSZ signal from resolved sources}
\label{subsec:catalogue}
\begin{figure*}
\includegraphics[width=\columnwidth]{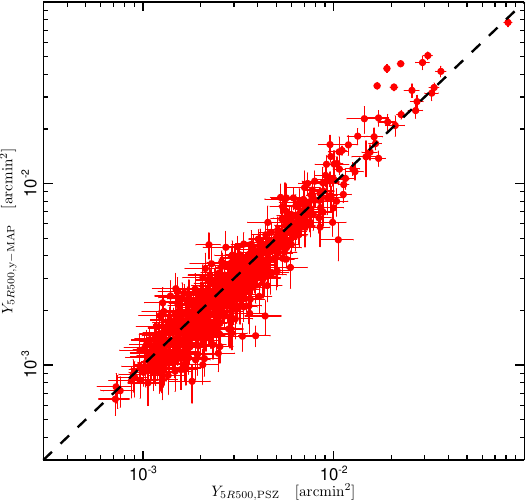}
\includegraphics[width=\columnwidth]{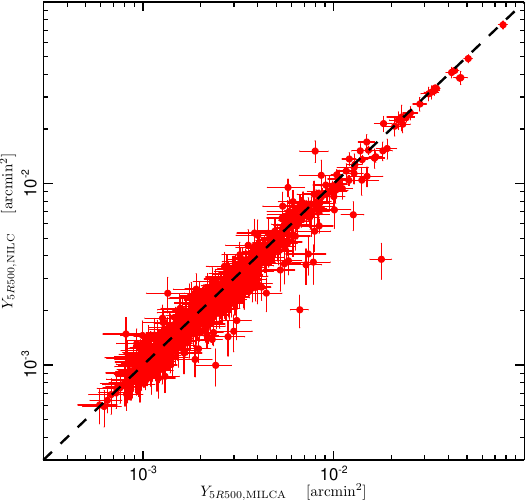}

\caption{Comparison of the measurements of $Y_{5R500}$. {\it Left}:
  the values derived from the detection methods used to build the \Planck\
  catalogue of clusters ($Y_{5R500,\mathrm{PSZ}}$), plotted against those
  from the all-sky reconstructed MILCA tSZ map ($Y_{5R500,y\mathrm{-MAP}}$).
  {\it Right}: the {\tt MILCA} ($Y_{5R500,\mathrm{MILCA}}$) versus {\tt NILC}
  ($Y_{5R500,\mathrm{NILC}}$) all-sky tSZ effect maps. The equality
  relationship is marked as a dashed black line. A least-squares
  bisector method fit to the data lead to slopes of $1.09\pm 0.02$
  and $1.08\pm0.02$ for the MILCA 
  and NILC Compton parameter maps, respectively. \label{fig:catcomp}}
\end{figure*}

As a first validation step of the Compton parameter maps we perform a
blind search for the SZ signal coming from resolved sources and compare it to
the \Planck\ catalogue of SZ sources \citep{planck2013-p05a}.  The latter
comprises 861
confirmed clusters out of 1227 cluster candidates and 54 {\sc class1}
highly reliable candidate clusters.

\subsubsection{Yields}

Two lists of SZ sources above a signal-to-noise ratio threshold of 4.5 are
constructed from both {\tt MILCA} and {\tt NILC} all-sky Compton
parameter maps outside a 33\% Galactic mask. The point source detections are
undertaken using two methods.

\begin{itemize}

\item {\tt SMATCH}, in which sources are detected using the {\tt
  SEXtractor} algorithm \citep{sextractor} over the whole sky divided
into 504 patches. A single frequency matched filter \citep{melin2006}
is then applied to measure the SZ flux density and signal-to-noise ratio
using the \citet{Arnaud2010} pressure profile. Using this method, we detect
843 and 872 sources in {\tt MILCA} and {\tt NILC}, respectively.

\item {\tt MHWS}, in which  SZ sources are detected in the maps using
{\tt IFCAMEX}
\citep[{\tt MHW2},][]{GonzalezNuevo:2006p2545,LopezCaniego:2006p2546}.
The flux density and signal-to-noise ratio are then estimated using {\tt SEXtractor}
on $3.65^{\circ} \times 3.65^{\circ}$ patches. We detect 1036 and 1740 sources
in {\tt MILCA} and {\tt NILC}, respectively, with this method.

\end{itemize}

The difference between the yields of the two 
methods is understandable, as {\tt SMATCH} is by construction
dedicated to the search for SZ sources and the precise measurement of
their flux (including assumptions on the spatial distribution of the
SZ signal), whereas {\tt MHWS} targets all types of compact source
(including IR and radio sources) and uses a more ``generic'' flux
estimation procedure.

We have compared these two lists of sources with 790 confirmed clusters
and {\sc class1} high reliability candidates from the \Planck\ catalogue
of SZ sources that fall outside the 33\% Galactic mask. The association is
performed on the basis of the source positions within a search radius
of 10$^{\prime}$ (the resolution of the SZ all-sky maps). We found 583
and 529 matches in the {\tt MILCA} source list with the SMATCH and
MHWS methods, respectively (614 and 414 from the {\tt NILC} source
list). This match of $52$ to $77$\% per cent, respectively.
This is consistent with the results in \citet{Melin:2012p2056}, which show that indirect
detection methods based on reconstructed $y$-maps are less efficient
at extracting clusters of galaxies than dedicated direct methods such
as those used to build the \Planck\ catalogue of SZ sources
\citep[i.e., {\tt MMF1}, {\tt MMF3} and {\tt PwS,}][]{2002MNRAS.336.1057H,2006A&A...459..341M,PwSII,planck2013-p05a}.

\subsubsection{Photometry}

Of more importance than a comparison of yields is the comparison in
terms of photometry. For all-sky map detections that are associated with clusters in the \Planck\ SZ catalogue, the SZ flux measurement from the all-sky maps correlates very well with the maximum likelihood value of the integrated Compton parameter, $Y_{5R_{500}}$\footnote{$R_{500}$ refers to the radius inside which
the mean density is 500 times the critical density at the cluster
redshift.}, provided by the dedicated SZ-detection methods in the \Planck\ SZ
catalogue. As shown in the left panel of Fig.~\ref{fig:catcomp}, the
correlation is very tight, with little dispersion (0.1 dex).
We note that the few points at high $Y_{5R500}$ that lie significantly above
the one-to-one line are not unexpected; they correspond to nearby and extended
clusters. On the one hand, the significance of SZ flux measurement increases
with the flux. On the other hand, the catalogue detection methods are not
optimized for the extraction of such
extended sources \citep[see][for details]{planck2013-p05a}.  Therefore
they tend to miss part of the SZ flux, which is recovered, together
with a better estimate of the cluster size, from the Compton parameter
map directly. 

As a sanity check, we have also matched the list of sources detected by a
given method using both {\tt MILCA} and {\tt NILC} maps in order to compare
the SZ photometry. The right panel of Fig.~\ref{fig:catcomp} shows very good
agreement between the methods. There is only $0.07$ and $0.01$ dex dispersion
between them for the {\tt SMATCH} and {\tt MHWS} extraction methods,
respectively.

Together, these results indicate that we can be confident in the fidelity with
which the tSZ signal is reconstructed over the whole sky by the {\tt MILCA}
and {\tt NILC} methods.

\section{Angular power spectrum of the reconstructed \textit{y}-map}
\label{sec:powerspec}

\begin{figure}
\includegraphics[width=\columnwidth]{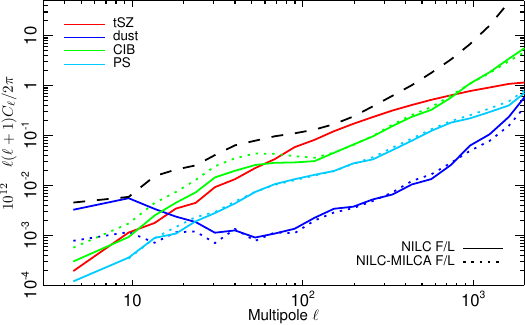}
\caption{Angular power spectrum of the main foreground contributions
  as estimated using the {\tt FFP6} simulations. We plot the
  diffuse Galactic emission (blue), clustered CIB (green) and point
  source (cyan) contributions, as well as the tSZ signal (red). The
  solid and dotted lines correspond to the {\tt NILC} F/L and to the
  {\tt NILC}-{\tt MILCA} F/L cross-power spectra, respectively.  For
  illustration we also show the \Planck\ instrumental noise auto-power
  spectrum (dashed black line) in the {\tt MILCA} Compton parameter
  map.
\label{fig:simu_pws}}
\end{figure}
 
\subsection{Methodology}

To estimate the power spectrum of the tSZ signal we use the {\tt
  XSPECT} method \citep{Tristram2005} initially developed for the
cross-correlation of independent detector maps.  {\tt XSPECT} uses
standard {\tt MASTER}-like techniques \citep{2002ApJ...567....2H} to
correct for the beam convolution and the pixelization, as well as the
mode-coupling induced by masking foreground contaminated sky
regions. 

We apply {\tt XSPECT} to the FIRST and LAST $y$-maps obtained using {\tt NILC} and {\tt MILCA}. We consider the following map pairs: the
{\tt MILCA} FIRST and LAST ({\tt MILCA} F/L); the {\tt NILC} FIRST and
LAST ({\tt NILC} F/L); and the {\tt NILC} FIRST and {\tt MILCA} LAST
({\tt NILC-MILCA} F/L), or equivalently the {\tt MILCA} FIRST and {\tt NILC}
LAST ({\tt MILCA-NILC} F/L).  As the noise is
uncorrelated between the map pairs the resulting power spectrum is
not biased and we preserve the variance. 

In the following, all the
spectra will use a common multipole binning scheme, which was defined in
order to minimize the correlation between adjacent bins at low
multipoles and to increase the signal-to-noise at high multipole
values. Error bars in the spectrum are computed analytically from the auto-power and cross-power spectra of the pairs of maps, as
described in \citet{Tristram2005}. All of our Compton parameter
maps assume a circular Gaussian beam of 10$^{\prime}$ FWHM.  The additional filtering at large angular scales in the {\tt MILCA} Compton parameter maps is
also accounted for and deconvolved.

 \begin{figure}
 \centering
 \includegraphics [width=\columnwidth]{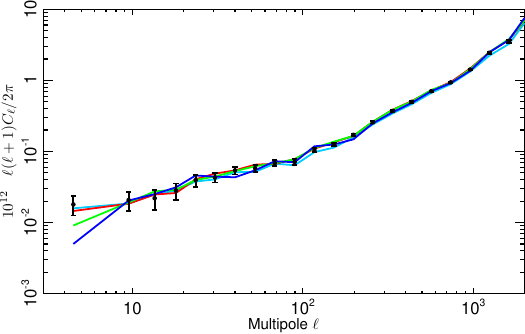}
\caption{Angular cross-power spectra of the \Planck\ {\tt NILC} F/L
reconstructed Compton parameter maps for different Galactic masks, removing
30\% (cyan), 40\% (black points and error bars), 50\% (red), 60\% (green),
and 70\% (blue) of the sky. \label{fig:galcut}}
\end{figure}

\subsection{Foreground contamination}
\label{subsec:forecont}

The challenge in computing the tSZ power spectrum
is to estimate and minimize foreground contamination. We do not intend here to provide a detailed foreground analysis, but rather to identify the main
foreground contaminants at different multipoles. We first
identify the dominant foregrounds in the reconstructed Compton
parameter maps. To do so, we apply to the {\tt FFP6} simulated
maps the linear combination weights of {\tt NILC} and {\tt MILCA}
derived from the real data. In this way we have
constructed maps of the expected foreground contamination in the final
Compton parameter maps. 

Figure~\ref{fig:simu_pws} shows the angular power spectra for these
reconstructed foreground contamination maps. We use the PSMASK and a
conservative common Galactic mask that leaves 50\% of the sky.
The Galactic mask is constructed by removing the 50\% brightest
regions of the sky in the 857\,GHz intensity map,
as detailed below in Sect.~\ref{sec:lowell}.  We show the diffuse
Galactic contamination (blue), the clustered CIB
contamination (green), and point source contamination
(cyan).  We consider here the foreground contamination in the
cross-power spectra of the {\tt NILC} F/L (dotted lines) and {\tt
  NILC}-{\tt MILCA} F/L maps (solid lines). The tSZ power spectrum
for the {\tt FFP6} simulations is plotted in red.  For
illustration we also show the \Planck\ instrumental noise power
spectrum (dashed black line) in the {\tt MILCA} Compton parameter map.
We clearly observe that, as expected, the diffuse Galactic emission
(mainly thermal dust), dominates the foreground contribution at low
multipoles.  For large multipoles the clustered CIB and point source
contributions dominate the power spectrum.  However, it is important to notice
that the tSZ signal dominates the angular power spectrum in the
multiple range $ 100 < \ell < 800$.  We also note that foreground
contamination differs depending on the reconstruction method, and we find
that {\tt MILCA} is more affected by foreground contamination. However, we
also find that at large angular scales the diffuse Galactic dust
contamination is significantly lower in the {\tt NILC}-{\tt MILCA} F/L
cross-power spectrum than in the {\tt NILC} F/L cross-power
spectrum. This indicates that the residual dust contamination is not
100 \% correlated between the reconstructed {\tt MILCA} and {\tt NILC}
Compton parameter maps.  In contrast, the clustered CIB and point
source contamination levels are similar for the two cross-power spectra at
high multipoles, indicating that the residual contamination is essentially
100\% correlated between the {\tt MILCA} and {\tt NILC} maps.

\subsubsection{Low-multipole contribution\label{sec:lowell}}

The diffuse Galactic foreground contribution can be significantly
reduced by choosing a more aggressive Galactic mask.  Assuming that at
large angular scales the Compton parameter maps are mainly affected by
 diffuse Galactic dust emission, we have tested several Galactic masks
by imposing flux cuts on the \Planck\ 857\,GHz channel intensity map.
In particular we investigated masking out 30\%, 40\%, 50\%, 60\%, and 70\% of
the sky.  The edges of these masks have been apodized to limit ringing
effects on the reconstruction of the angular power spectrum.
Figure~\ref{fig:galcut} presents the angular cross-power spectrum of
the reconstructed {\tt NILC} F/L Compton parameter maps for some of
these Galactic masks: 30\% (cyan); 40\% (black); 50\% (red); 60\% (green);
70\% (blue); and the PSMASK. We find that when masking 40\% or more of
the sky the tSZ angular power spectrum does not change
significantly. That is why, conservatively, we select the 50\% mask
(GALMASK50 hereafter), which will be used in the remainder of our
analysis.

\begin{figure}[t]
 \centering
 \includegraphics [width=\columnwidth]{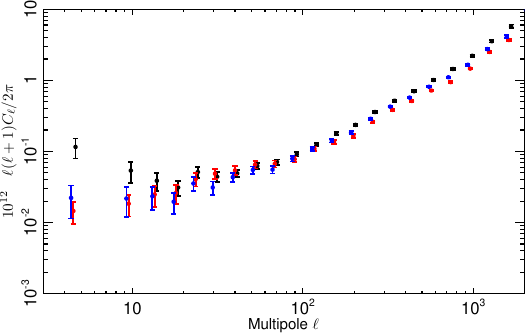}
\caption{Angular cross-power spectra between the reconstructed \Planck\
{\tt MILCA} F/L (black), {\tt NILC} F/L (red), and {\tt NILC}-{\tt MILCA}
F/L (blue) maps.\label{fig:nilcmilcafortest}}
 \end{figure}

\begin{figure}[t]
 \centering
\includegraphics [width=\columnwidth]{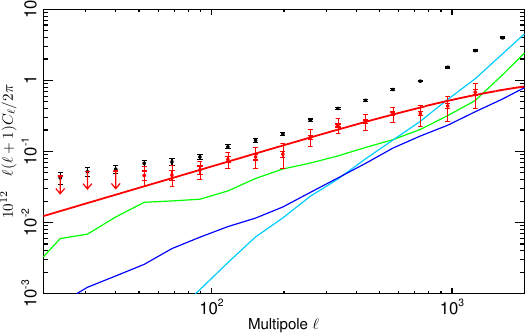}
\caption{{\tt NILC} F/L cross-power spectrum before (black points) and after
(red points) foreground correction, compared to the power spectra of the
physically motivated foreground models.  Specifically we show: clustered CIB
(green line); infrared sources (cyan line); and radio sources (blue line). 
The statistical (thick line) and  total (statistical plus foreground, thin
line), uncertainties are also shown.  Additionally we
show the best-fit tSZ power spectrum model presented in
Sect.~\ref{powerspectcosmo} as a solid red line.
 \label{fig:refinedforegrounds}}
 \end{figure}

We checked if the foreground contribution in the reconstructed
\Planck\ Compton parameter maps also depends on the reconstruction
method.  From the analysis of the {\tt FFP6} simulations we have
found that the contribution from foregrounds in the {\tt NILC} and
{\tt MILCA} Compton parameter maps is not the same, and it is not
fully correlated.  Similar results are found for the \Planck\ data.
Figure~\ref{fig:nilcmilcafortest} shows the cross-power spectra
between the {\tt MILCA} F/L maps (black)\footnote{The excess of power
  at low $\ell$ observed in the {\tt MILCA} F/L maps angular
  cross-power spectrum is due to the deconvolution from the extra
  low-multipole filtering in the {\tt MILCA} maps, discussed in
  Sect. \ref{subsec:compsep}}, the {\tt NILC} F/L maps (red) and the
{\tt NILC}- {\tt MILCA} F/L maps (blue), as a function of $\ell$. We observe
that the {\tt MILCA} F/L cross-power spectrum shows a larger amplitude than
the {\tt NILC} F/L cross-power spectrum.  This is most probably due to a
larger foreground contamination in the {\tt MILCA} Compton parameter map.

In addition, we find that the {\tt NILC}-{\tt MILCA} F/L\footnote{And 
equivalently {\tt MILCA}-{\tt NILC} F/L that is not shown in the Figure.} 
cross-power spectrum shows the lowest amplitude at low multipoles
($\ell < 100$). This is due to a reduction of the dust contamination
in the cross-correlation of the {\tt NILC} and {\tt MILCA} Compton
parameter maps with respect to the dust contamination in the original
maps. We also find that the {\tt NILC}-{\tt MILCA} F/L lies between the
{\tt MILCA} F/L and {\tt NILC} F-L cross-power spectra at high multipoles.
This can be explained by the differences in the clustered CIB contamination in
the {\tt MILCA} and {\tt NILC} Compton parameter maps. An accurate model of
the clustered CIB power spectrum is available. However, this is not the case
for the dust
contamination power spectrum, and thus we restrict the power spectrum
analysis presented in Sect.~\ref{powerspectcosmo} to $\ell > 60$.

Hereafter, we will consider the {\tt NILC} F/L cross-power spectrum as a
baseline for cosmological analysis, with the {\tt NILC}-{\tt MILCA} F/L
cross-power spectrum being used to cross-check the results.

\subsubsection{High-multipole contribution}
\label{refinedforegroundmodel}

The high-$\ell$ contamination from clustered CIB and point
sources affects the measurement of the tSZ spectrum and its
cosmological interpretation. Realistic models fitted to the
\Planck\ data are thus needed.  We take advantage of the
capability of \Planck\ to measure and constrain these
foreground emissions and use the outputs of
\citet{planck2011-6.6} and \citet{planck2013-pip56} for the clustered CIB
modelling.  For the six \Planck\  HFI frequencies considered in this
paper, the clustered CIB model consists of six auto-power spectra
and 24 cross-power spectra.  For frequencies above 217\,GHz, these
spectra are fitted in \citet{planck2013-pip56} to the measured CIB,
consistently with \citet{planck2011-6.6}. The model is extrapolated
at 100 and 143\,GHz following~\citet{2012ApJ...757L..23B} and
\citet{planck2011-6.6}. The uncertainties in the clustered-CIB
model are mainly due to the 
cross-correlation coefficients that relate the cross-power spectra to
the auto-power spectra. Following~\citet{planck2013-pip56} we
consider 5\% global uncertainties on those coefficients.  

We use the \citet{2012ApJ...757L..23B} model to compute the star-forming dusty
galaxy contribution.  Finally, we use the \citet{2011A&A...533A..57T} model, 
fitted to the \Planck\ ERCSC~\citep{planck2012-VII}, for extragalactic radio
sources. Notice that these models are also used for the study of the
clustered CIB with \Planck\ \citep{planck2013-pip56}. 

We now estimate the residual power spectrum in the $y$-map after
component separation. We apply the {\tt MILCA} or {\tt
  NILC} weights to Gaussian-realization maps drawn using the cross-
and auto-spectra of each component at the six \Planck\  HFI
frequencies. The residual power spectrum in the $y$-map can also be
estimated in the spherical harmonic domain, as detailed in
Appendix~\ref{app:forecont}. We have tested the consistency between
the two approaches and we give here results for a map-based estimate
using a total of 50 all-sky simulations for each of the foreground
components. Specific simulations, varying the foreground models, were 
also performed to propagate the 5\% global uncertainties of the
model-coefficients (which include the overall
uncertainties in the CIB modelling)
into the estimated residual power spectrum. We find
a 50\% uncertainty in the amplitude of each residual spectrum
(clustered CIB, star-forming dusty galaxies, and radio sources) in the
$y$-map.

Figure~\ref{fig:refinedforegrounds} shows the {\tt NILC} F/L cross-power
spectrum before (black points) and after (red
points) foreground correction, using the refined foreground models
presented above. We also show the clustered CIB (green), infrared
source (cyan), and radio source (blue) power spectrum contributions.

 \begin{figure}
 \centering
 \includegraphics [width=\columnwidth]{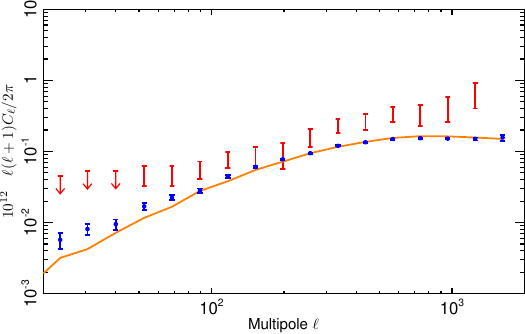}
\caption{Comparison of the tSZ angular power spectrum estimated from 
  the cross-power-spectrum of the {\tt NILC} F/L maps (black) with the
  expected angular power spectrum of the confirmed clusters in the
  \Planck\ Cluster Sample (orange line).  In red we plot the
  {\tt NILC} F/L cross-power spectrum after masking these
  clusters. The green points correspond to the difference of these two
  cross-power spectra. The cross-power spectrum between the {\tt NILC}
  Compton parameter map and the simulated detected cluster map is
  shown in blue. \label{fig:conta_cib}}
 \end{figure}

\subsection{Contribution of resolved clusters to the tSZ power spectrum}
\label{subsec:resolvedclusters}

We simulate the expected Compton parameter map for the detected and
confirmed clusters of galaxies in the
\Planck\ catalogue~\citep{planck2013-p05a} from their measured
integrated Compton parameter, $Y_{\mathrm{5R500}}$. The orange solid line in
Fig.~\ref{fig:conta_cib} shows the power spectrum of this simulated map.
Figure~\ref{fig:conta_cib} also shows the cross-power spectrum of the
{\tt NILC} F/L maps (in black). In red we plot the cross-power
spectrum of the {\tt NILC} F/L maps after masking the
confirmed clusters from the PSZ catalogue. The green curve corresponds
to the difference of the two cross-power spectra, with and without
masking the clusters. It is in good agreement with the modelled power
spectrum of the confirmed clusters of galaxies. We also compute the
cross-power spectrum of the simulated cluster map and the
\Planck\ reconstructed Compton parameter {\tt NILC} map. This is shown
in blue in the figure. Here again, the signal is consistent with the
expected power spectrum of the confirmed \Planck\ clusters of
galaxies.  

These results show that a significant fraction of the
signal in the reconstructed \Planck\ Compton parameter maps is due to
the tSZ effect of detected and confirmed clusters of
galaxies, verifying the SZ nature of the signal.  In addition, by
comparing the tSZ power spectrum from the resolved clusters with the
marginalized tSZ power spectrum presented in Sect.~\ref{sec:cosmo}, we
deduce that the measured tSZ spectrum includes an additional tSZ
contribution from unresolved clusters and diffuse hot gas.

\section{Analysis of high-order statistics}
\label{sec:higorderstat}

The power spectrum analysis presented above only provides 
information on the 2-point statistics of the Compton parameter
distribution over the sky. An extended characterization of the field can be
performed by studying the higher-order moments in the 1D PDF of the
map, or by measuring 3-point statistics, i.e., the bispectrum.

\begin{figure}
\includegraphics[width=\columnwidth]{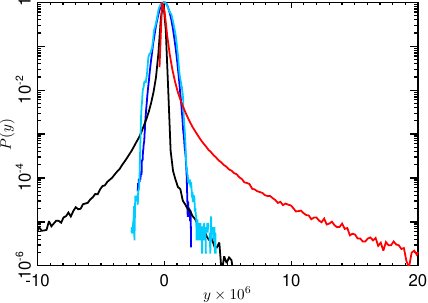}
\caption{1D PDF for the {\tt FFP6} simulation maps considering the {\tt MILCA}
linear combination weights obtained for the real data.
The tSZ effect (red), diffuse Galactic emission (cyan), clustered CIB (blue),
and radio source (black) contributions to the 1D PDF are shown.
\label{fig:py_simu}}
\end{figure}

\begin{figure}
\includegraphics[width=\columnwidth]{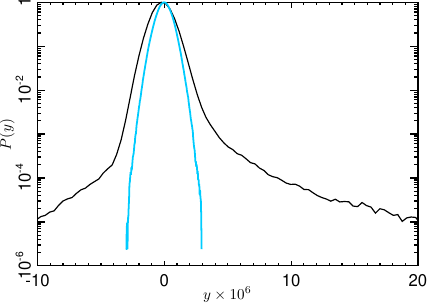}
\caption{1D PDF of the \Planck\ $y$-map (black) and of the null map (cyan)
for the {\tt MILCA} method.
\label{fig:py_data}}
\end{figure}
\subsection{1D PDF analysis}
\label{methods:1D PDF}

We performed an analysis of the 1D PDF of the {\tt NILC} and {\tt
  MILCA} reconstructed Compton parameter maps. For the tSZ effect we
expect an asymmetric distribution with a significantly positive tail
\citep{RubinoMartin:2003p1790}. We thus focus on the asymmetry of the
distribution and its unnormalized skewness.  First, we filter the maps
in order to enhance the tSZ signal with
respect to foreground contamination and noise.  To avoid residual point source ringing effects near the edges of the combined PSMASK and GALMASK50 masks we apodize them. We follow the
  approach of \cite{Wilson:2012p2102} and use a filter in harmonic
space, constructed from the ratio between the angular power spectrum of
the expected tSZ signal in the {\tt FFP6} simulations and the
power spectrum of the null $y$ maps.  We smooth this ratio using a
21-point square kernel and normalize it to one by dividing by its
maximum value.  Notice that this filter only selects the multipole
range for which the tSZ signal is large with respect to the noise, and
thus, it does not modify the non-Gaussianity properties.  Furthermore,
we have found that the filter used here behaves better than the more
traditionally used Wiener filter, as it is less affected by
point-source ringing. Following this procedure, the 1D PDF of the filtered
Compton parameter map, $P(y)$, is computed from the histogram of the pixels.

Figure~\ref{fig:py_simu} shows the 1D PDF for the {\tt FFP6}
simulation maps combined using the weights of the {\tt MILCA} linear
combination of the real data. We present in red the 1D PDF of the tSZ effect, 
which is clearly asymmetric, with a positive tail as expected.
Moreover, the asymptotic slope of this red curve at high values of $y$
scales almost as $P(y) \propto y^{-2.5}$, implying that the underlying
source counts should scale in the same way (i.e., $dn/dy \propto
y^{-2.5}$). This is the predicted scaling behaviour for clusters
\citep[e.g.,][]{1995A&A...300..335D,RubinoMartin:2003p1790}, and
indeed, it is the scaling that we find in the actual number counts of
clusters in the simulation used.  Similarly, the 1D PDF for radio
sources (black) is also asymmetric, but with a negative tail. By
contrast, the clustered CIB (blue) and diffuse Galactic emission
(cyan) distributions are symmetric to first approximation.  From this
analysis we see that, as expected, the filtering enhances the tSZ
effect with respect to foregrounds and therefore helps in their
discrimination.

For illustration, Fig.~\ref{fig:py_data} shows the 1D PDF for
the {\tt MILCA} Compton parameter map in black. This is the
convolution of the 1D PDF of the different components in the map: the tSZ
effect; foregrounds; and noise. Indeed, it clearly shows three
distinct contributions: a Gaussian central part that exceeds
slightly the contribution from noise, as expected from the null map 1D
PDF (cyan curve); a small negative tail, corresponding most likely to
residual radio sources; and a positive tail corresponding mainly to
the tSZ signal.  A direct computation of the slope of the full $P(y)$
function in Fig.~\ref{fig:py_data} shows that it converges to $-2.5$
for $y > 10^{-5}$, as predicted from the cluster counts.

A simple analysis of the measured 1D PDF can be performed by considering
the asymmetry of the distribution:
\begin{equation}
A\equiv \int_{y_{\mathrm{p}}}^{+\infty}P(y)dy - \int^{y_{\mathrm{p}}}_{-\infty}P(y)dy,
\end{equation}
where $y_{\mathrm{p}}$ is the peak value of the normalized
distribution ($\int P(y)dy = 1$).  In addition, the non-Gaussianity of
the positive tail can be quantified by
\begin{equation}
\Delta =  \int_{y_{\mathrm{p}}}^{+\infty}\left[P(y)-G(y)\right]dy,
\end{equation}
with $G(y)$ the expected distribution if fluctuations were only due to
noise.  For the {\tt NILC} Compton parameter map we find $A=0.185$ and
$\Delta=0.065$.  Equivalently, for the {\tt MILCA} Compton parameter
map we find $A=0.26$ and $\Delta=0.11$.  These results are consistent
with a positive tail in the 1D PDF, as expected for the tSZ effect.
The differences between the {\tt NILC} and {\tt MILCA} results come
mainly from the difference in filtering. Similar values are obtained
for the FFP6 simulations, with $A=0.12$ and $\Delta=0.05$ for NILC
and $A=0.30$ and $\Delta=0.13$ for MILCA.

Alternatively, we can also compute the skewness of the obtained
distribution, $\int y^{3} P(y)dy/\left(\int y^{2} P(y)dy
\right)^{3/2}$.  Following \cite{Wilson:2012p2102} we have chosen here
a hybrid approach, by computing the unnormalized skewness of the
filtered Compton parameter maps outside the 50\% sky mask. In particular we
have computed the
skewness of the \Planck\ data Compton parameter maps
$\langle y^{3}\rangle$, and of
the null maps $\langle y_{\mathrm{NULL}}^{3}\rangle$.  For the {\tt FFP6}
simulations, we computed these for the tSZ
component $\langle y_{\mathrm{FFP6,SZ}}^{3}\rangle$ and for the sum of
all astrophysical components $\langle y_{\mathrm{FFP6,ALL}}^{3}\rangle$.
Table~\ref{table:skewness} shows the
results for the {\tt NILC} and {\tt MILCA} maps. The different filtering
function derived for the {\tt NILC} and {\tt MILCA} $y$-maps prevents a direct
one-to-one comparison of the skewness values. However, the comparison of
each map with the FFP6 simulations of the tSZ component and of the sum
of all components clearly shows that the contribution of foregrounds is minor
in both maps, and suggests that the measured skewness is
mainly dominated
by the tSZ signal, as one would expect from Figs.~\ref{fig:py_simu}
and \ref{fig:py_data}. By comparing the measured and model skewness, we
present constraints on $\sigma_{8}$
in Sect.~\ref{subsec:cosmohighorder}.

\subsection{Bispectrum}
\label{bispectrumm}

\begin{table}[tmb]
\begingroup
\newdimen\tblskip \tblskip=5pt
\caption{Unnormalized skewness, multiplied by $10^{18}$. \label{table:skewness}}                          
\nointerlineskip
\vskip -1mm
\footnotesize
\setbox\tablebox=\vbox{
   \newdimen\digitwidth 
   \setbox0=\hbox{\rm 0} 
   \digitwidth=\wd0 
   \catcode`*=\active 
   \def*{\kern\digitwidth}
   \newdimen\dpwidth 
   \setbox0=\hbox{.} 
   \dpwidth=\wd0 
   \catcode`!=\active 
   \def!{\kern\dpwidth}
\halign{\hbox to 1.5cm{#\leaderfil}\tabskip 1em&
     \hfil#\hfil \tabskip 1em&
     \hfil#\hfil \tabskip 1em&
     \hfil#\hfil \tabskip 1em&
     \hfil#\hfil \tabskip 0em \cr
\noalign{\doubleline}
\omit Method& $\left<y^{3}\right>$& $\left<y_{\mathrm{NULL}}^{3}\right>$&
 $\left<y_{\mathrm{FFP6,SZ}}^{3}\right>$& $\left<y_{\mathrm{FFP6,ALL}}^{3}\right>$\cr
\noalign{\vskip 3pt\hrule\vskip 5pt}
{\tt NILC}&  $1.78$& $-0.0001$& $2.17$& $2.09$\cr
{\tt MILCA}& $1.50$& $*0.0004$& $1.46$& $1.21$\cr
\noalign{\vskip 3pt\hrule\vskip 5pt}
}
}
\endPlancktable 
\endgroup
\end{table}

Since the SZ signal is non-Gaussian, significant statistical information is  contained in the bispectrum, complementary to the power spectrum
\citep{RubinoMartin:2003p1790,Bhattacharya:2012p2458}. We therefore
compute the bispectrum of the {\tt NILC} and {\tt MILCA} reconstructed
Compton parameter maps. The results presented here use the binned
bispectrum estimator described in \citet{Bucher2010} and
\citet{Lacasa2012}, which is also used for the \Planck\ primordial
non-Gaussianity analysis \citep{planck2013-p09a}.  We mask the maps
with the combined PSMASK and GALMASK50, remove the best-fit monopole
and dipole outside the mask, and degrade the resolution to
$N_{\mathrm{side}}=1024$ to reduce computing time. We use a multipole
bin size $\Delta \ell = 64$ and a maximum multipole $\ell_\mathrm{max}
= 2048$ for the analysis. To correct for the bias introduced by
masking, we have produced non-Gaussian simulations with a tSZ-like
bispectrum and we have convolved the simulated maps with a Gaussian
beam of 10$^{\prime}$ FWHM. We compute the bispectrum of the simulated
full-sky and masked maps and measure the average ratio between the
two. This ratio is used to correct the measured bispectra and flag
unreliable $(\ell_1,\ell_2,\ell_3)$ configurations, for which mask
effects are too large to be corrected.

\begin{figure}[t]
\begin{center}
\includegraphics[width=4.3cm]{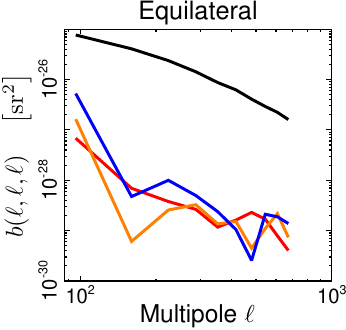}
\includegraphics[width=4.3cm]{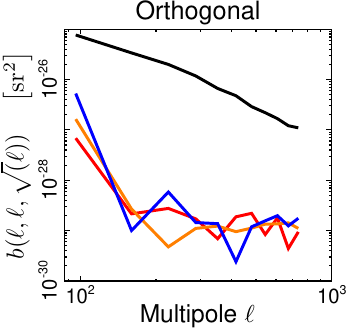}
\includegraphics[width=4.3cm]{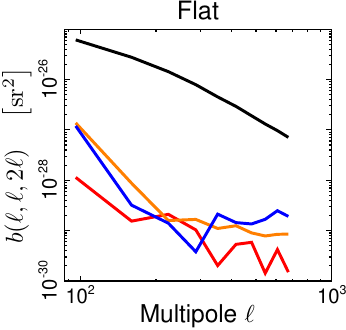}
\includegraphics[width=4.3cm]{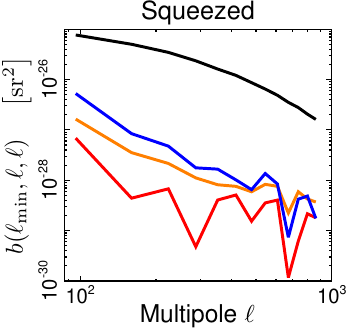}
\caption{Binned bispectra of the {\tt FFP6} tSZ map, and foreground residuals for the {\tt MILCA} component separation. The black line represents the tSZ bispectrum and the red line the clustered CIB. In addition, we plot the bispectrum for the Galactic diffuse free-free (orange), and the thermal dust (dark blue) emission. \label{Fig:bispectre_milca_residuals}}
\end{center}
\end{figure}

\begin{figure}[t]
\begin{center}
\includegraphics[width=4.3cm]{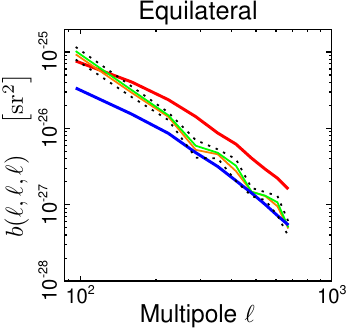}
\includegraphics[width=4.3cm]{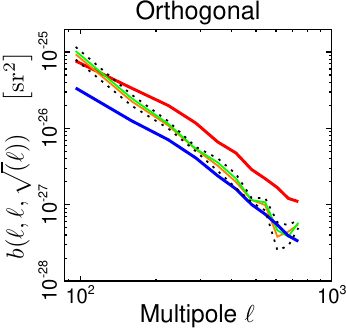}
\includegraphics[width=4.3cm]{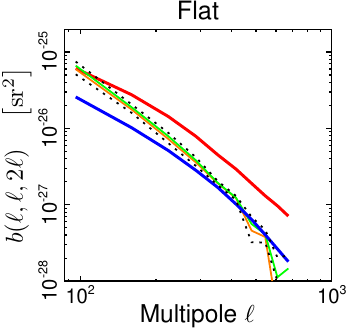}
\includegraphics[width=4.3cm]{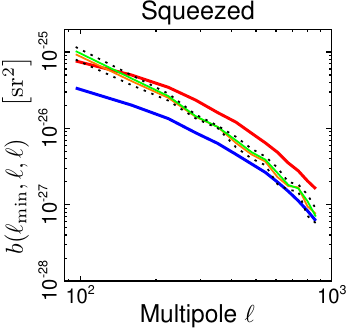}
\caption{tSZ measured bispectra for the {\tt MILCA} (green) and {\tt NILC}
(orange) Compton parameter maps, compared with the {\tt FFP6} tSZ bispectrum
(in red) and with the projected map of the catalogue of detected clusters
(in blue).  Uncertainties on the measured tSZ power bispectra are given by
the dotted lines. \label{Fig:bispectre_DX9Rec_catalog_ffp6}}
\end{center}
\end{figure}

We checked that foreground residuals do not significantly affect the
recovered tSZ bispectrum by using the {\tt FFP6} simulations
described previously. In the case of the {\tt MILCA} reconstructed map
(more affected by foregrounds), for example,
Fig.~\ref{Fig:bispectre_milca_residuals} shows the tSZ bispectrum as
well as the (absolute value of the) bispectra of the different foreground
residuals.  This is shown for some some special configurations, namely
equilateral $(\ell,\ell,\ell)$, orthogonal isosceles
$(\ell,\ell,\sqrt{2}\ell)$, flat isosceles $(\ell,\ell,2\ell)$ and
squeezed $(\ell_\mathrm{min},\ell,\ell)$.  The foreground residuals
yield negligible bispectra, at least one order of magnitude smaller
than the tSZ bispectrum over the multipoles of interest.

In Fig.~\ref{Fig:bispectre_DX9Rec_catalog_ffp6} we compare the tSZ
bispectrum measured on \Planck\ data, with the tSZ bispectrum of the
{\tt FFP6} simulation and with the bispectrum of the maps of
detected clusters in the \Planck\ catalogue presented above. Clusters
from the \Planck\ catalogue contribute an important fraction of the
measured bispectrum, at least 30\% on large angular scales and more on
smaller angular scales; the bispectrum therefore also probes the
unresolved tSZ signal, as was the case for the power spectrum. On large
angular scales this may be the signature of the clustering of less
massive dark matter halos inside the large-scale
structures. Alternatively large angular scales may be affected by
foreground residuals.

\section{Cosmological Interpretation}
\label{sec:cosmo}

\begin{table}[tmb]
\begingroup
\newdimen\tblskip \tblskip=5pt
\caption{Marginalized bandpowers of the angular power spectrum of the \Planck\
tSZ Compton parameter map (in dimensionless $(\Delta T/T)^2$ units),
statistical and foreground errors, and
best-fit tSZ power spectrum and number counts models
(also dimensionless).\label{tab:bandpowers}}                          
\nointerlineskip
\vskip -1mm
\footnotesize
\setbox\tablebox=\vbox{
   \newdimen\digitwidth 
   \setbox0=\hbox{\rm 0} 
   \digitwidth=\wd0 
   \catcode`*=\active 
   \def*{\kern\digitwidth}
   \newdimen\dpwidth 
   \setbox0=\hbox{.} 
   \dpwidth=\wd0 
   \catcode`!=\active 
   \def!{\kern\dpwidth}
\halign{\tabskip 0em\hfil#\hfil\tabskip 1em&
     \hfil#\hfil \tabskip 1em&
     \hfil#\hfil \tabskip 1em&
     \hfil#\hfil \tabskip 1em&
     \hfil#\hfil \tabskip 1em&
     \hfil#\hfil \tabskip 1em&
     \hfil#\hfil \tabskip 0em \cr
\noalign{\doubleline}
*$\ell_{\mathrm{min}}$& *$\ell_{\mathrm{max}}$& *$\ell_{\mathrm{eff}}$&
 ${\ell(\ell+1)C_{\ell}/2\pi}$& $\sigma_{\mathrm{stat}}$&
 $\sigma_{\mathrm{fg}}$&  Best-fit\cr
\noalign{\vskip 5pt}
\omit& & & $[10^{12} {y}^2]$& $[10^{12} {y}^2]$& $[10^{12} {y}^2]$&
 $[10^{12} {y}^2]$\cr
\noalign{\vskip 3pt\hrule\vskip 5pt}
**21& **27& $**23.5$&  $<0.045$& $ \dots$& $ \dots$& $0.014$\cr
\noalign{\vskip 2pt}
**27& **35& $**30.5$&  $<0.052$& $ \dots$& $ \dots$& $0.019$\cr
\noalign{\vskip 2pt}
**35& **46& $**40!*$&     $<0.053$& $ \dots$& $ \dots$& $0.025$\cr
\noalign{\vskip 2pt}
**46& **60& $**52!*$&    $**0.046$& $0.007$& $^{+0.014}_{-0.011}$& $0.032$\cr
\noalign{\vskip 2pt}
**60& **78& $**68!*$&    $**0.047$& $0.007$& $^{+0.015}_{-0.012}$& $0.042$\cr
\noalign{\vskip 2pt}
**78& *102& $**89!*$&    $**0.056$& $0.007$& $^{+0.015}_{-0.013}$& $0.055$\cr
\noalign{\vskip 2pt}
*102& *133& $*117!*$&    $**0.077$& $0.008$& $^{+0.020}_{-0.016}$& $0.072$\cr
\noalign{\vskip 2pt}
*133& *173& $*152!*$&    $**0.084$& $0.008$& $^{+0.029}_{-0.025}$& $0.094$\cr
\noalign{\vskip 2pt}
*173& *224& $*198!*$&    $**0.092$& $0.009$& $^{+0.040}_{-0.033}$& $0.121$\cr
\noalign{\vskip 2pt}
*224& *292& $*257!*$&    $**0.158$& $0.009$& $^{+0.046}_{-0.040}$& $0.157$\cr
\noalign{\vskip 2pt}
*292& *380& $*335!*$&    $**0.232$& $0.012$& $^{+0.056}_{-0.050}$& $0.203$\cr
\noalign{\vskip 2pt}
*380& *494& $*436!*$&    $**0.264$& $0.013$& $^{+0.069}_{-0.064}$& $0.261$\cr
\noalign{\vskip 2pt}
*494& *642& $*567!*$&    $**0.341$& $0.017$& $^{+0.080}_{-0.081}$& $0.332$\cr
\noalign{\vskip 2pt}
*642& *835& $*738!*$&    $**0.340$& $0.024$& $^{+0.102}_{-0.110}$& $0.417$\cr
\noalign{\vskip 2pt}
*835& 1085& $*959!*$&    $**0.436$& $0.035$& $^{+0.149}_{-0.171}$& $0.515$\cr
\noalign{\vskip 2pt}
1085& 1411& $1247!*$&    $**0.681$& $0.059$& $^{+0.222}_{-0.272}$& $0.623$\cr
\noalign{\vskip 3pt\hrule\vskip 5pt}
}
}
\endPlancktable 
\endgroup
\end{table}

\subsection{Power spectrum analysis}
\label{powerspectcosmo}

As a measure of structure growth, the tSZ power spectrum can provide
independent constraints on cosmological parameters and potentially improve their
precision. As shown by \citet{2002MNRAS.336.1256K}, the power spectrum
of the tSZ effect is highly sensitive to the normalization of the
matter power spectrum, commonly parameterized by the rms of the $z =
0$ mass distribution on $8\, h^{-1}\,\mathrm{Mpc}$ scales, $\sigma_{8}$,
and to the total amount of matter
$\Omega_\mathrm{m}$. We expect the tSZ power spectrum to
also be sensitive to other cosmological parameters, e.g., $\Omega_\mathrm{b}$,
$H_{0}$, and $n_{\mathrm{s}}$. For reasonable external priors
on those parameters, however, the variations are
expected to be negligible with respect to those introduced by changes in
$\Omega_\mathrm{m}$ and $\sigma_{8}$ and are not considered here.
Finally, we also expect the tSZ power spectrum amplitude to be sensitive to the
``mass bias'', $b$.  A full joint analysis cosmological parameters and mass bias
is not possible with the current data and so we have chosen here to fix
the mass bias to $b=0.2$ following results in the companion
Planck paper on cosmological constraints from Planck SZ cluster counts
\citep{planck2013-p15}. 
Note that final cosmological constraints depend on this choice.
\begin{figure}[]
\begin{center}
\includegraphics[width=1.1\columnwidth]{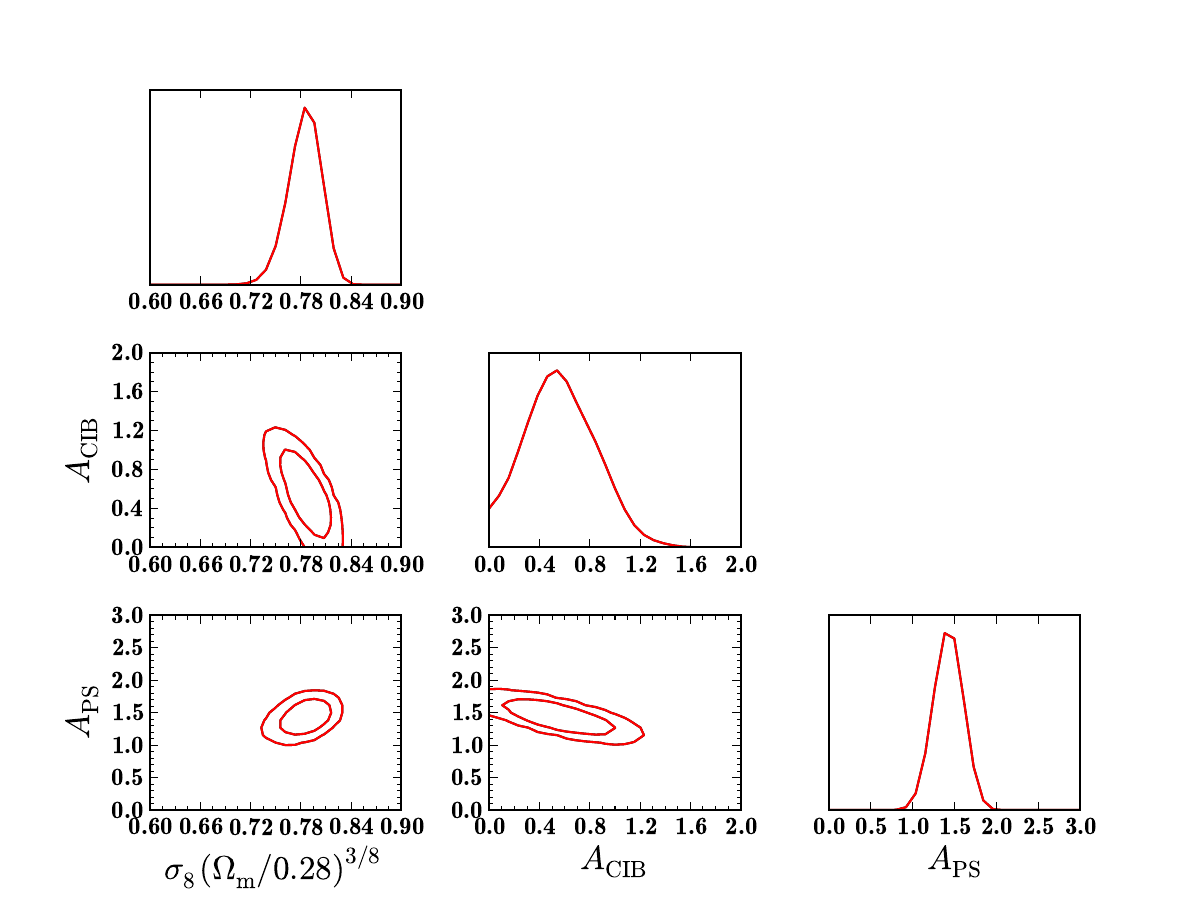}
\end{center}
\caption{2D and 1D likelihood distributions for the combination of
 cosmological parameters $\sigma_{8} (\Omega_{\mathrm{m}}/0.28)^{0.40}$,
 and for the foreground parameters $A_{\mathrm{CIB}}$ and $A_{\mathrm{PS}}$.
 We show the 68.3\% and 95.4\% C.L. contours (in orange).
\label{fig:compfitall}}
\end{figure}

\begin{figure}
\centering
\includegraphics[width=\columnwidth]{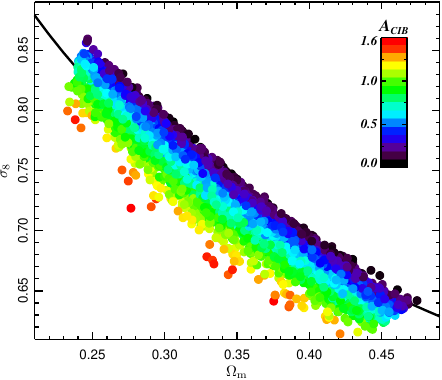}
\caption{Likelihood samples derived from the MCMC chains. The points
 represent pairs of values of $\Omega_{\mathrm{m}}$ and
 $\sigma_8$. Only values within the 95.4\% C.L. contours are shown.
 The clustered CIB amplitude is colour-coded according to
 the value of $A_{\mathrm{CIB}}$, from low (blue) to high (red). The
 black solid line shows the theoretical degeneracy between the two
 cosmological parameters.}
\label{fig:points}
\end{figure}

Cosmological constraints are obtained from a fit of the {\tt NILC} F/L
cross-power spectrum, for the 50\% mask, assuming a three-component
model: tSZ; clustered CIB; and radio and infrared point sources. For
$\ell > 60$, we can reasonably neglect the Galactic dust
contamination. For $\ell > 1411$ the total signal in the tSZ map is
dominated by noise. We thus restrict our analysis to the multipole
range $60 < \ell < 1411$.  The measured power spectrum,
$C_{\ell}^{\mathrm{m}}$, is modelled as:
\begin{equation}
C_{\ell}^{\mathrm{m}} = C_{\ell}^{\mathrm{tSZ}} (\Omega_{\mathrm{m}}, \sigma_{8}) + A_{\mathrm{CIB}} \ C_{\ell}^{\mathrm{CIB}} + A_{\mathrm{PS}} \  (C_{\ell}^{\mathrm{IR}} + C_{\ell}^{\mathrm{Rad}}). 
\end{equation}
Here $C_{\ell}^{\mathrm{tSZ}} (\Omega_{\mathrm{m}},\sigma_{8})$ is the
tSZ power spectrum, $C_{\ell}^{\mathrm{CIB}}$ is the clustered CIB
power spectrum, and $C_{\ell}^{\mathrm{IR}}$ and
$C_{\ell}^{\mathrm{Rad}}$ are the infrared and radio source power
spectra, respectively.

Following Eq.~(\ref{twohalomodel}), the tSZ spectrum is computed using
the 2-halo model, the \citet{Tinker:2008p1782} mass function, and the
\citet{Arnaud2010} universal pressure profile. In particular, we use
the numerical implementation presented
in~\citet{Taburet2009,Taburet2010a,Taburet11}, and integrating in
redshift from 0 to 3 and in mass from $10^{13}\,\mathrm{M}_{\odot}$ to
$5 \times 10^{15}\,\mathrm{M}_{\odot}$. Our model allows us to compute the
tSZ power spectrum at the largest angular scales. It is consistent with the tSZ
spectrum presented in~\citet{2012MNRAS.423.2492E}, which was used as a
template in the CMB cosmological analysis in
\citet{planck2013-p08} and \citet{planck2013-p11}.

\begin{figure*}[]
\begin{center}
\includegraphics[width=1.9\columnwidth,height=12cm]{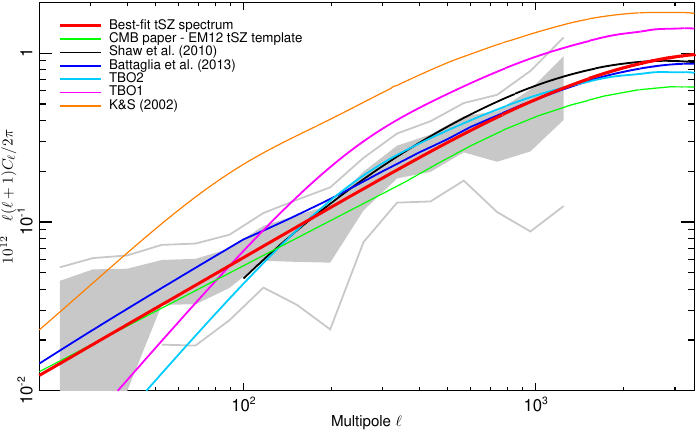}
\end{center}
\caption{Marginalized bandpowers of the \Planck\ tSZ power spectrum
 with total (statistical plus foreground) uncertainties (red
 points). The red solid line represents the best-fit tSZ power
 spectrum model. We also show as a blue solid line the best-ft tSZ
 power spectrum obtained from the analysis of cluster number
 counts~\citep{planck2013-p15}. The tSZ power spectrum template used
 in the CMB cosmological
 analysis~\citep{planck2013-p08,planck2013-p11} is presented as a
 green solid line. For comparison, we also show the SPT \citep[][orange diamond]{2012ApJ...755...70R} and
ACT \citep[][cyan diamond]{Sievers:2013p2161} constraints on the tSZ
power spectrum at $\ell = 3000$.
\label{fig:comp:final}}
\end{figure*}

Foreground contamination is modelled following
Sect.~\ref{refinedforegroundmodel}. 
As discussed there, the main uncertainties in the
residual power spectrum translate into up to 50\% uncertainty in the
clustered CIB and point source amplitudes. We thus allow for a
variation of the normalization amplitudes for the clustered CIB,
$A_{\mathrm{CIB}}$, and for the point sources, $A_{\mathrm{PS}}$, with
Gaussian priors centred on 1 with standard deviation 0.5.

We have not considered explicitly the expected correlation
between the tSZ effect and the CIB.
However, using the formalism in~\citet{Addison:2012p2790}, we have
performed simulations of the expected effect and find that to a
reasonable level of approximation the shape of the tSZ and clustered
CIB cross-power spectrum is very similar to that of the clustered CIB
power spectrum. Therefore, in our simplified modelling, the clustered
CIB normalization factor, $A_{\mathrm{CIB}}$, also accounts for this
component. 

We assume a Gaussian approximation for the likelihood
function. Best-fit values and uncertainties are obtained using an
adapted version of the {\tt Cosmo-MC} algorithm~\citep{PhysRevD.66.103511}.
Only $\sigma_{8}$ and $\Omega_{\mathrm{m}}$ are allowed to vary
here. All other cosmological parameters are fixed to their best-fit
values as obtained in Table~2 of ~\citet{planck2013-p11}. The
normalization amplitudes, $A_{\mathrm{CIB}}$ and $A_{\mathrm{PS}}$,
considered as nuisance parameters, are allowed to vary between 0 and
3. For the range of multipoles considered here, the tSZ angular power
spectrum varies like $C_\ell \propto \sigma_8^{8.1}
\Omega_\mathrm{m}^{3.2}$. The results are thus presented in terms of
this parameter combination.

Figure~\ref{fig:compfitall} presents the 2D and 1D likelihood
distributions for the cosmological parameter combination $\sigma_{8}
\Omega_{\mathrm{m}}^{3.2/8.1}$, or equivalently
$\sigma_8\Omega_{\mathrm{m}}^{0.40}$ and for the foreground nuisance
parameters.  The best-fit values and error bars for each parameter are
given by $\sigma_{8} (\Omega_{\mathrm{m}}/0.28)^{0.40}=0.784 \pm 0.016$,
$\sigma_8=0.74 \pm 0.06$, $\Omega_\mathrm{m}=0.33 \pm 0.06$,
$A_{\mathrm{CIB}}=0.55 \pm 0.26$, and $A_{\mathrm{PS}}=0.14 \pm 0.13$.
It is worth noting that these values are obtained
in a specific framework, all other cosmological parameters being fixed and a
fiducial fixed model used for the signals. 
Relaxing this framework would likely weaken the constraints presented
in this paper.

Figure~\ref{fig:points} shows the
degeneracy between the two cosmological parameters from the Monte Carlo
Markov chains (MCMC),
as well as the theoretical degeneracy (solid black line). It
also shows the dependency on $A_{\mathrm{CIB}}$ (colour coded from low
values in blue to high values in red).  While the combination
$\sigma_{8} (\Omega_{\mathrm{m}}/0.28)^{0.40}$ is well determined,
marginalized constraints on $\sigma_8$ and $\Omega_\mathrm{m}$ are
weaker.  To check the robustness of our results, we performed the same
cosmological analysis using the {\tt NILC}-{\tt MILCA} F/L cross-power
spectrum presented in Fig.~\ref{fig:nilcmilcafortest}.
Although the foreground level is
different, we find compatible results at the 1$\,\sigma$
level. Furthermore, our constraints are in
good agreement with those derived from the \Planck\ cluster number
count analysis \citep{planck2013-p15}, which shows a similar
$\sigma_8$--$\Omega_\mathrm{m}$ degeneracy line. Conversely, our
findings exhibit some tension with the constraints derived from the
\Planck\ primary CMB analysis \citep{planck2013-p11}, which finds larger values
of $\sigma_8$ and $\Omega_\mathrm{m}$.  However, as discussed in
\cite{planck2013-p15}, the constraints from the SZ signal depend
significantly on the assumed value of the mass bias.

The red points in Fig.~\ref{fig:comp:final} correspond to the
marginalized \Planck\ tSZ power spectrum (from the {\tt NILC} F/L
cross-power spectrum), compared to the best-fit theoretical model
presented above (solid red line). Foreground uncertainties are derived
from the likelihood curves of the nuisance parameters and added in
quadrature to the statistical uncertainties, providing the total
errors plotted here.  Table~\ref{tab:bandpowers} presents the \Planck\
marginalized tSZ power spectrum, together with statistical and foreground
uncertainties, and the best-fit tSZ power spectrum model. 
In the range $\ell=60$--1411, the \Planck\ tSZ power spectrum can be
approximated by a power law of the form
\begin{equation}
\ell(\ell+1)C_{\ell}/2\pi =
(1.0 \pm 0.2) \times  10^{-15 }\ell^{(0.91\pm0.03)}\,.
\end{equation} 
 
The measured tSZ power spectrum is in remarkable agreement with the
tSZ power spectrum (blue solid line) computed using the cluster count
best-fit parameters \citep{planck2013-p15}. We also show in
Fig.~\ref{fig:comp:final} (green line) the tSZ template used in the
\Planck\ CMB analysis \cite{planck2013-p11}. This template is
renormalized by a simple scaling factor
using the best-fit $\sigma_{8} (\Omega_{\mathrm{m}}/0.28)^{0.40}$
The difference in shapes of the two spectra is due to the
different assumptions used for the scaling relation between SZ signal and
mass \cite[see,][]{2012MNRAS.423.2492E}.
We also show the SPT \citep[][orange diamond]{2012ApJ...755...70R} and
ACT \citep[][cyan diamond]{Sievers:2013p2161} constraints on the tSZ
power spectrum at $\ell = 3000$, which are consistent with our best-fit
model within $\pm2\sigma$ and illustrate that the tSZ spectrum starts to
turn over at higher $\ell$.

\begin{figure}[t]
\begin{center}
\includegraphics[width=0.975\columnwidth]{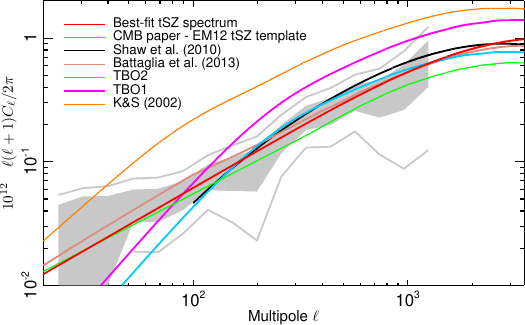}
\end{center}
\caption{Comparison of the \Planck\ tSZ power spectrum and best-fit model
 with existing models in the literature. The \Planck\ tSZ power spectrum 
 and the $\pm1$ and 2$\,\sigma$ error bars are shown in grey.
 We also show the \Planck\ tSZ
 power spectrum best-fit models derived in this paper (red) and from
 the analysis of cluster number
 counts~\citep[][blue]{planck2013-p15}. The tSZ power spectrum
 template used in the CMB cosmological
 analysis~\citep[][green]{planck2013-p08,planck2013-p11} is also
 shown.  We additionally show the tSZ power spectrum models from
 hydrodynamic simulations \citep[][brown]{Battaglia:2012p1841},
 from $N$-body simulations plus semi-analytical dust gas models
 \citep[][cyan; TBO1 and purple; TBO2]{Trac:2011p1795}, and from analytical calculations
 (\citealt[][black]{Shaw2010}; \citealt[][orange]{Komatsu:2002p1799}).
\label{fig:comp:models}}
\end{figure}

In Fig.~\ref{fig:comp:models}, we compare the \Planck\ tSZ measurements
of the power spectrum to a set of predicted spectra. We consider the
predictions derived from hydrodynamical simulations
\citep[][brown]{Battaglia2010,Battaglia:2012p1841}, from $N$-body
simulations plus semi-analytical models
\citep[][purple and cyan]{Trac:2011p1795} and from analytical calculations
(\citealt[][black]{Shaw2010}; \citealt[][orange]{Komatsu:2002p1799}).
These models were computed originally for the set of
cosmological parameters in~\citet{Hinshaw:2012p2598} with
$\sigma_{8}=0.8$ and have been rescaled in amplitude to our best-fit value
for $\sigma_8^{8.1} \Omega_\mathrm{m}^{3.2}$. We note that there is some
dispersion in the predicted amplitudes and shapes of the tSZ power spectrum.
These differences reflect the range of methodologies and assumptions
used both in the physical properties of clusters and in the
technical details of the computation. The latter includes
differences in the redshift ranges and also in the
mass intervals probed by the limited sizes of the simulation boxes of the
hydrodynamical simulations. Analytical predictions are also sensitive to
the model ingredients, such as the mass function, mass bias and scaling
relations adopted.

We see from Fig.~\ref{fig:comp:models} that most of the models presented
above (the tSZ template for CMB analyses, plus
the \citealt{Battaglia:2012p1841}, \citealt{Shaw2010} and TBO2 models)
provide reasonable fits to the data,
while the others (TBO1 and \citealt{Komatsu:2002p1799})
are clearly not consistent.
The TBO1 model was a highly simplified approach superseded by
TBO2 \citep{Trac:2011p1795}.  The \citet{Komatsu:2002p1799}
prediction shows a significantly different shape compared with all the other
models; this is not well understood and we will not consider it further.
We have performed a simplified likelihood analysis to evaluate the
uncertainties in cosmological parameters induced
by the uncertainties in the modelling of the cluster physics. 
We replace our own model of the tSZ power spectrum by the models discussed
above (excluding the TBO1 and \citealt{Komatsu:2002p1799} ones) and recompute
$\sigma_8(\Omega_{\mathrm{m}}/0.28)^{0.40}$, $A_{\mathrm{CIB}}$, and
$A_{\mathrm{PS}}$ from a simple linear fit to the {\tt NILC} F/L
cross-power spectrum. 
We obtain values for $\sigma_8(\Omega_{\mathrm{m}}/0.28)^{0.40}$
between 0.768 and 0.798,
which lie within the 1$\,\sigma$ uncertainties (0.016) presented above. 

The \Planck\ data allow us for the first time to probe the
large and intermediate angular scales ($\ell=46$ to $\ell=1085$) for the tSZ spectrum;
furthermore, and, as
shown in \citet{planck2013-p05a}, \Planck\ is particularly sensitive to
the SZ signal from massive clusters not probed by other
experiments. The \Planck\ tSZ measurement will hence permit us to better understand the integrated tSZ
contribution of the whole population of clusters, including resolved
and nearby clusters, the correlated SZ signal, and possible diffuse hot
gas.

\subsection{High-order statistics}
\label{subsec:cosmohighorder}

The estimates of tSZ non-Gaussianity, e.g., the unnormalized skewness and
bispectrum, are very sensitive to $\sigma_8$. Using the models
presented in Sect.~\ref{sec:theory} we can show that the unnormalized
skewness of the tSZ fluctuation, $\langle T^{3}(\mathbf{n})\rangle$
scales approximately as $\sigma_{8}^{11}$, whereas the
amplitude of the bispectrum scales as $\sigma_8^\alpha$ with
$\alpha=11$--$12$, as shown by \cite{Bhattacharya:2012p2458}.
We do not consider in the following the dependency of
the bispectrum and the unnormalized skewness on other cosmological
parameters, since all such dependencies are expected to be significantly lower
than for $\sigma_{8}$ \citep{Bhattacharya:2012p2458}.

We derive constraints on $\sigma_{8}$ by comparing the measured
unnormalized skewness and bispectrum amplitudes with those obtained from
simulations of the tSZ effect. This approach is strongly limited by
systematic uncertainties and the details of the theoretical modelling \cite[see][]{Hill:2013p2067}.
 
From the measured unnormalized skewness of the filtered {\tt MILCA}
and {\tt NILC} Compton parameter maps discussed in Sect.~\ref{methods:1D PDF}
and by comparing them to the value measured in the {\tt FFP6}
simulations we can derive constraints on $\sigma_{8}$.  Uncertainties due
to foreground contamination are computed using the {\tt FFP6}
simulations and are accounted for in the final error bars.  The tSZ
component of the {\tt FFP6} simulations was obtained from a hybrid
simulation including a hydrodynamic component for $z<0.3$ plus extra
individual clusters at $z>0.3$, and with $\sigma_{8} = 0.789$. Using
these simulations we obtain $\sigma_{8} = 0.775$ for {\tt NILC} and
$\sigma_{8} = 0.783$ for {\tt MILCA}. Combining the two results and
considering model and foreground uncertainties we obtain $\sigma_{8} =
0.779\pm 0.015 (68\% \ \mathrm{C.L.})$. Notice that
the uncertainties are mainly dominated by foreground contamination.
Model uncertainties here only account for the expected dependence of the
unnormalized skewness upon $\sigma_{8}$, as shown in
Sect.~\ref{sec:theory}. We have neglected, as was also the case
in~\citet{Wilson:2012p2102}, the dependence on other cosmological
parameters. We have also not considered any uncertainties coming from the
combination of the hydrodynamical and individual cluster
simulations. Because of these constraints, our error bars might be
underestimated.

The comparison of the measured bispectrum obtained from
the \Planck\ Compton parameter maps with the {\tt FFP6} simulation
tSZ bispectrum shows an offset of about a factor of two on small
angular scales, $300<\ell<700$, which we attribute to the differences
in cosmological parameters. Using the scaling of the bispectrum with
$\sigma_8$, its uncertainty, as well as the uncertainty on the
bispectra ratio, we obtain $\sigma_8=0.74\pm0.04 (68\%\ \mathrm{C.L.})$.
As was the case for the unnormalized skewness, we
neglected here the dependence on other cosmological parameters and the
uncertainties in the {\tt FFP6} simulations.  Thus the error
bar might again be somewhat underestimated. However, we expect those
additional uncertainties to be smaller than the error bars we quote.

\section{Conclusion}
\label{conclusions}

Because of its wide frequency coverage from 30 to 857\,GHz, the
\Planck\ satellite mission is particularly well suited for the
measurement of the thermal Sunyaev-Zeldovich effect.
Working with the \Planck\ frequency channel
maps from 100 to 857\,GHz, we have reconstructed the tSZ signal over the full
sky using tailored component separation methods. In this paper, we have analysed
the first all-sky tSZ map quantified in terms of the Compton parameter and
with an angular resolution of 10\arcmin. 

We have characterized the reconstructed \Planck\ all-sky 
Compton parameter map in terms of blind detection of tSZ sources,
and the angular power spectrum and higher order statistics via the study
of its 1D PDF and bispectrum. In all cases we have identified, characterized
and carefully modelled the contamination by foreground emission.
This is mainly due to diffuse Galactic thermal dust emission at large angular scales ($\ell \la 60$), and clustered CIB and Poisson-distributed radio and
infrared sources at smaller angular scales (dominating at $\ell \ga 500$). 
Diffuse Galactic thermal dust emission is tackled via a conservative masking of 
the brightest 50\% of the sky in the \Planck\ 857\,GHz channel map.
The CIB and point-source contamination are modelled in a way which is
consistent with the findings of \citet{planck2011-6.6} and
\citet{planck2013-pip56}.

We have produced the first measurement of the SZ power spectrum on
large angular scales, ranging over
$0.17^{\circ} \la \theta \la 3.0^{\circ}$.  In this range, the tSZ power
spectrum is almost insensitive to the physics of cluster cores. The
detected tSZ signal likely arises from the contribution of warm and
hot diffuse gas distributed within groups and clusters, sampling the
whole halo mass function, as well as within the larger-scale
filamentary structures.

We have modelled the tSZ power spectrum via a halo-model analytical
approach, in order investigate its dependence on $\sigma_{8}$
and $\Omega_{\mathrm{m}}$ and to test it against the measured \Planck\
tSZ power spectrum.  Moreover, we performed an analysis of the 1D PDF
and bispectrum of the \Planck\ $y$-map to infer independent
constraints. We find, in the present framework, that the best-fit
normalization parameter $\sigma_8$ from the three independent analysis
ranges between $(0.74 \pm 0.06)$ and $(0.78 \pm 0.02)$ at 68\% C.L.
for the high-order statistics and power spectrum analyses, respectively.

These values are lower than those derived from analysis of primary
CMB anisotropies \citep{planck2013-p11}. 
More refined analysis and modelling will be needed to understand this
difference, since the tension may have several possible origins.
Some of the difference may be due to
specific choices in the tSZ modelling, e.g., the mass bias
\citep[see][for a detailed discussion on its effect of its effect on cluster
counts]{planck2013-p15}.  Other differences could arise from the foreground
modelling, in particular at high frequencies, above 217\,GHz. 

The observed consistency between constraints derived from the cluster
number counts in \citet{planck2013-p15} and from the present work
provides a coherent view of the gas content in halos and in
larger-scale structures. As such, this \Planck\ tSZ
measurement constitutes the first step towards building a comprehensive
understanding of the integrated tSZ effect due to cosmic structure on
all scales and at all density contrasts.

\begin{acknowledgements}
The development of \Planck\ has been supported by: ESA; CNES and
CNRS/INSU-IN2P3-INP (France); ASI, CNR, and INAF (Italy); NASA and DoE
(USA); STFC and UKSA (UK); CSIC, MICINN, JA and RES (Spain); Tekes,
AoF and CSC (Finland); DLR and MPG (Germany); CSA (Canada); DTU Space
(Denmark); SER/SSO (Switzerland); RCN (Norway); SFI (Ireland);
FCT/MCTES (Portugal); and PRACE (EU). A description of the \Planck\
Collaboration and a list of its members, including the technical or
scientific activities in which they have been involved, can be found
at \url{http://www.sciops.esa.int/index.php?project=Planck&page=Planck_Collaboration}.
We acknowledge the use of the HEALPix software.
\end{acknowledgements}

\bibliographystyle{aa}

\bibliography{szCls.bib,Planck_bib.bib}

\appendix

\section{Foreground contamination in the final tSZ power spectrum}
\label{app:forecont}
Since we are using modified Internal Linear Combination methods to estimate the final \Planck\ Compton parameter map we can write it as
\begin{equation}
\hat{y} (\theta,\phi) = \sum_{\nu} \sum_{b} W_{\nu}^{b} (\theta,\phi) \left(  F^{b} (\theta,\phi) * M_{\nu}  (\theta,\phi)  \right),
\label{ydefinition}
\end{equation}
where $M_{\nu}  (\theta,\phi)$  is the \Planck\ map for frequency channel
$\nu$, $F^{b} (\theta,\phi)$ is a circular filtering function
for the multipole interval $b$, and $ W_{\nu}^{b} (\theta,\phi)$ are the
weights of the internal linear combination into that multipole range.
Decomposing $\hat{y} (\theta,\phi)$ in spherical harmonics we obtain 
\begin{equation}
\hat{y}_{\ell,m} = \sum_{\nu} \sum_{b} {W_{\nu}^{b}}_{\ell,m;\ell^{\prime},m^{\prime}}   F^{b}_{\ell^{\prime}}  * {M_{\nu}}_{\ell^{\prime},m^{\prime}}.
\end{equation}

Then using spherical harmonic convolution properties \citep[see for example][]{Tristram2005} and assuming overlap in the multipole range selected by the
filter functions, $F^{b}_{\ell}$, then the power spectrum is given by
\begin{equation}
C_{\ell}^{y,y} = \sum_{b} \sum_{b^{\prime}} \sum_{\nu} \sum_{\nu^{\prime}} \sum_{\ell^{\prime}} \cal{M}_{\ell,\ell^{\prime}}^{W_{\nu}^{b},W_{\nu^{\prime}}^{b^{\prime}}} F^{b}_{\ell^{\prime}} F^{b^{\prime}}_{\ell^{\prime}}  C_{\ell^{\prime}}^{M_{\nu},M_{\nu^{\prime}}},
\label{cldefinition}
\end{equation}
where
$\cal{M}_{\ell,\ell^{\prime}}^{W_{\nu}^{b},W_{\nu^{\prime}}^{b^{\prime}}}$
represents the mode-coupling matrix associated with $W_{\nu}^{b}
\times W_{\nu^{\prime}}^{b^{\prime}}$.

For each \Planck\ channel the sky signal can be expressed as the sum of
multiple components, including CMB, tSZ, diffuse Galactic emission, radio and
IR point sources, and clustered CIB, such
that the \Planck\ Compton parameter is given by
\begin{equation}
\hat{y} = y + y^{\mathrm{CMB}} + \sum_{c} y^{c},
\end{equation}
where $c$ sums over the different foreground contributions. 
By construction $y_{\mathrm{CMB}} = 0$ and thus, assuming no correlation between foreground components, the estimated tSZ spectrum can be expressed as
\begin{equation}
C^{\hat{y},\hat{y}}_{\ell} = C^{\mathrm{tSZ}}_{\ell}
 + \sum_{c} C^{y^{c},y^{c}}_{\ell}.
\end{equation}

\noindent Using Eq.~(\ref{ydefinition}) we write
\begin{equation}
y^{c} (\theta,\phi) = \sum_{\nu} \sum_{b} W_{\nu}^{b} (\theta,\phi)
 \left(  F^{b} (\theta,\phi) * M^{c}_{\nu}  (\theta,\phi)  \right),
\end{equation}
and thus, using Eq.~(\ref{cldefinition}), we have
\begin{equation}
C^{y^{c},y^{c}}_{\ell} = \sum_{b}  \sum_{b^{\prime}} \sum_{\nu}
 \sum_{\nu^{\prime}}
 \sum_{\ell^{\prime}} M_{\ell,\ell^{\prime}}^{W_{\nu}^{b},W_{\nu^{\prime}}^{b}}
 F^{b}_{\ell^{\prime}} F^{b}_{\ell^{\prime}}
 C_{\ell^{\prime}}^{M^{c}_{\nu},M^{c}_{\nu^{\prime}}}.
\end{equation}

The latter expression can be simplified assuming a common spatial distribution
of the foreground emission across frequencies and a well defined spectral
energy density, $f^{c}_{\nu}$, so that it reads
\begin{equation}
C^{y^{c},y^{c}}_{\ell} = \sum_{b} \sum_{\nu} \sum_{\nu^{\prime}}
 \sum_{\ell^{\prime}}
 \cal{M}_{\ell,\ell^{\prime}}^{W_{\nu}^{b},W_{\nu^{\prime}}^{b^{\prime}}}
 F^{b}_{\ell^{\prime}} F^{b^{\prime}}_{\ell^{\prime}} \
 f^{c}_{\nu} f^{c}_{\nu^{\prime}} \  C_{\ell^{\prime}}^{M^{c},M^{c}}.
\end{equation}

\noindent Let us now look at the cross-correlation between the estimated
Compton parameter map and a particular sky
component at one of the observation frequencies
$M_c^{\nu'}(\theta,\phi) = \sum_{\ell m}T_{c;\ell m}^\nu
 Y_{\ell m}(\theta,\phi)$. We define the cross-power spectrum as
\begin{equation}
C_\ell^{\hat{y},c}(\nu) =
 \frac{1}{2\ell+1}\sum_{m=-\ell}^{\ell} \hat{y}_{\ell m} M^{\nu *}_{c;\ell m}
\end{equation}
and the statistical expectation of this quantity reads
\begin{eqnarray}
\left\langle C_\ell^{\hat{y},c}(\nu) \right\rangle &=&
 \frac{1}{2\ell+1}\sum_{m=-\ell}^{\ell}
\sum_{b,\nu'}\sum_{\ell'm'} W^{b \nu'}_{\ell\ell';mm'} F^b_{\ell'}
 \left\langle M^{\nu'}_{\ell'm'}M^{\nu *}_{c,\ell m}\right\rangle \\
&=& \frac{1}{2\ell+1}\sum_{m=-\ell}^{\ell}
 \sum_{b,\nu'}\sum_{\ell'm'} W^{b \nu'}_{\ell\ell';mm'} F^b_{\ell'}
 C_{c;\ell}^{\nu\nu'}\delta_{\ell\ell'}\delta_{mm'}\\
&=& \frac{1}{2\ell+1}\sum_{m=-\ell}^{\ell}
 \sum_{b,\nu'} W^{b \nu'}_{\ell\ell;mm} F^b_{\ell}C_{c;\ell}^{\nu\nu'}.
\end{eqnarray}
Assuming that we have a measure of $C_{c;\ell}^{\nu\nu'}$ and a way to derive
$W^{b \nu'}_{\ell\ell;mm}F^b_\ell$, we can compare the measured
cross-correlation $\hat{y}\times c$ to its theoretical expectation and thus
have a consistency check on each component's contribution to $\hat{y}$.

\raggedright

\end{document}